\documentclass[fleqn,usenatbib]{mnras}

\usepackage{bm}

\usepackage{newtxtext,newtxmath}
\usepackage[T1]{fontenc}
\usepackage{tabularx}
\DeclareRobustCommand{\VAN}[3]{#2}
\let\VANthebibliography\thebibliography
\def\thebibliography{\DeclareRobustCommand{\VAN}[3]{##3}\VANthebibliography}

\usepackage{subfigure}

\usepackage[dvipsnames]{xcolor} % for textcolor colors - SG

\usepackage{graphicx}	% Including figure files
\usepackage{amsmath}	% Advanced maths commands
\usepackage{multicol}
\title[Impact of stellar rotation on black holes]{The effect of stellar rotation on black hole mass and spin}

\author[S. Ghodla \& J.J. Eldridge]{Sohan Ghodla$^{\thanks{sohanghodla9@gmail.com}{1}}$,  J. J. Eldridge$^{1}$ \\
$^{1}$Department of Physics, University of Auckland, Private Bag 92019, Auckland, New Zealand}   

\date{Accepted XXX. Received YYY; in original form ZZZ}

\pubyear{2024}

\begin{document}
\label{firstpage}
\pagerange{\pageref{firstpage}--\pageref{lastpage}}
\maketitle


\begin{abstract}
   The gravitational wave signature of a binary black hole (BBH) merger is dependent on its component mass and spin.  If such black holes originate from rapidly rotating progenitors, the large angular momentum reserve in the star could drive a collapsar-like supernova explosion, hence substantially impacting these characteristics of the black holes in the binary.
   To examine the effect of stellar rotation on the resulting black hole mass and spin, we conduct a 1D general relativistic study of the end phase of the stellar collapse. 
   We find that the resulting black hole mass at times differs significantly from the previously assumed values. We quantify the dependence of the black hole spin magnitude on the hydrodynamics of the accretion flow, providing analytical relations for calculating the mass and spin based on the progenitor's pre-collapse properties. Depending on the nature of the accretion flow, our findings have implications for the black hole upper mass gap resulting from pair-instability supernovae, the maximum mass of a maximally rotating stellar black hole, and the maximum effective spin of a BBH formed in a tidally locked helium star - black hole binary.
\end{abstract}


%%%%%%%%%%%%%%%%%%%%%%%%%%%%%%%%%%%%%%%%%%%%%%%%%%

%%%%%%%%%%%%%%%%% BODY OF PAPER %%%%%%%%%%%%%%%%%%

\section{Introduction}

The black hole masses resulting from rapidly rotating pre-collapse stars are not well described by the pre-existing prescriptions that estimate the remnant mass produced at core-collapse \citep[e.g.][]{Fryer_Kalogera_2001, Eldridge_Tout_2004, Fryer2012}.
The most widely used remnant mass functions (resulting from the supernova explosion) are based on the work of \cite{Fryer2012}. They offer two variations of supernova models to calculate the resulting compact remnant mass, namely the \emph{Delayed} and \emph{Rapid} explosion mechanisms.
Both models rely on convection-enhanced neutrino heating (\citealt{Herant:1994}; see \citealt{Fryer:2021} for a recent review) to drive a successful explosion. The primary difference between them lies in the instability that could drive such an explosion. The Delayed mechanism is driven by a standing accretion shock instability \citep{blondin2003stability} and can produce an explosion with a delay of up to one second after the bounce, while the Rapid mechanism utilizes the Rayleigh-Taylor instability (e.g., \citealt{sharp1984overview}), and the explosion occurs within 0.25s.
Under a failed explosion, the Rapid mechanism leads to a direct collapse, while in the Delayed mechanism, some form of explosion may still be possible.

On the other hand, for progenitor stars with a substantial reservoir of angular momentum at core collapse, the supernova explosion might not be determined by convection and therefore may lead to remnant masses different from those predicted by the convection paradigm. This is because the angular momentum reservoir acts as an additional means of centrifugal support and the subsequent viscous dynamics offer an efficient mechanism for energy dissipation during the episode of the explosion.

\cite{Fryer2012} assumed that under the scenario of substantial rotation, the occurrence of relativistic jets during core collapse (e.g., \citealt{Woosley1993_collapsar, Woosley_1999}) would likely disrupt most of the star and only $\approx$ 5$M_\odot$ black hole would be produced. However, a number of subsequent studies, although at times working in the paradigm of rapidly rotating progenitor stars, have either continued to use the \cite{Fryer2012} remnant mass function (e.g., \citealt{Bavera_2020, Riley_2021}) or assumed a more or less direct collapse - i.e., a major mass fraction of the pre-core-collapse star ending up in the black hole, with some (typically the envelope) mass lost during the explosion (e.g., \citealt{deMink_Mandel2016, Marchant2016, Qin_2018, du_Buisson2020}). 
This is primarily due to the lack of a mass function that captures the effect of stellar rotation on the resulting black hole masses. 

Another quality of astrophysical black holes is their spin angular momentum, which, together with their mass, fully determines their property \citep{Israel:1967_nohair, Carter:1971_nohair, Bekenstein_72_nohair}. Although 1D studies exist that take into account the spin evolution of the black hole during collapse of the star (e.g., \citealt{Batta_2019, Bavera_2020, Fuller_2022}), they either do not account for the effect of wind outflows (e.g., \citealt{Bavera_2020}), a strong magnetic field, and/or relativistic dynamics on the nature of the accretion flow \citep{Batta_2019, Bavera_2020, Fuller_2022}. In this paper, we aim to address these shortcomings.

The collapse of a sufficiently rapidly rotating star should happen via the formation of an equatorial accretion disk. 
Such a supernova explosion paradigm, consisting of a newly formed spinning black hole penetrated by a strong magnetic field of the surrounding accretion disk, is also known as a collapsar \citep{Woosley1993_collapsar, Woosley_1999} and is believed to be primarily responsible for the observed long-duration $\gamma$-ray bursts (GRBs).
The dynamics of the accretion flow under a collapsar scenario has been extensively studied in the past (e.g., \citealt{Gammie:2004, Lee:2006, Kumar_2008, Tchekhovskoy_2010, Tchekhovskoy_2011, 2014_Sadowski, 2016_Sadowski, Fujibayashi_2020, Fujibayashi:2023, Murguia_Berthier:2020, Gottlieb:2023, Jacquemin_Ide:2023}).
The disk evolves viscously via magnetohydrodynamics, resulting in mass outflow and a transient energetic electromagnetic outburst. The resulting dynamics could be highly non-linear, and a detailed study would require a full 3D magnetohydrodynamic simulation (e.g., \citealt{skadowski2016three}). However, conducting such simulations is computationally expensive and time-consuming. Moreover, presently, most stellar models at core collapse are already 1D in nature, and thus, using these models to conduct a 3D study becomes less effective.

For the current work, we use a simpler approach to study the collapse dynamic, which is based on the earlier work of \cite{Kumar_2008} - and later numerically implemented in \cite{Fuller_2022}, which is a 1D non-relativistic, height-integrated, time-averaged formalism, but takes into account the effect of vertical wind outflows. We extend their work to a general relativistic framework, hence allowing us to more accurately consider the influence on the accretion flow in the inner region of the accretion disk around the newly formed Kerr black hole. We adopt two variable approaches to treat the accretion flow in the near-horizon region, namely the \cite{Novikov_Thorne_1973} accretion disk formalism and the magnetically arrested accretion disk formalism \citep{Tchekhovskoy_2011}.

We use this framework in conjunction with the \textsc{Mesa} generated stellar models to calculate the resulting black hole mass and spin relations that would be suitable for black holes born from rotating stars. For the latter, we provide fitting functions that can be readily used by future studies (in particular population synthesis) where the pre-collapse nature of the progenitor star is known. We then compare our results with the earlier work of \cite{Fryer2012}, highlighting the crucial difference between the two. We also discuss some implications of our work for the expected birth properties of stellar black holes.

The remainder of this paper is organized as follows. In Section~\ref{sec: numerical methods}, we discuss details of the \textsc{Mesa} models used in our analysis. Here, we also develop the formalism used for treating the collapsar dynamics. In Section~\ref{sec: results}, compare the relativistic formalism with the non-relativitic one. Here, we also present the mass and spin functions for black holes born from rotating stars and discuss their domain of validity. Then, in Section~\ref{sec: discussion}, we present some implications of our results, followed by a brief conclusion in Section~\ref{sec: conclusion}.

\vspace{-10pt}
\section{Method} \label{sec: numerical methods}

\subsection{Models for rotating stars} \label{sec: MESA models}

We use \textsc{Mesa} \citep{MESA2011, MESA2013, MESA2015, MESA2018, MESA2019} to generate models of variably rotating helium stars in the mass range 5-80$M_{\odot}$ (separated by $2M_\odot$) with their normalized initial angular velocity $\Omega_{\rm in}~\in~[0.01-0.92]$ (separated by 0.04), where the normalization is w.r.t. the critical angular velocity. These models are then evolved from the onset of helium burning to the core carbon depletion\footnote{The inlists for the \textsc{Mesa} model, the \textsc{Sage} manifolds notebook containing parts of the analytical calculation, and the \textsc{Python} script consisting of the numerical implementation of the formalism discussed in Section~\ref{sec: formalism} can be found in the data availability statement.}. No substantial change in the relevant structure (or the angular momentum reservoir) of the star is expected past core carbon depletion, as is evident from the output of such evolved \textsc{Mesa} models (e.g., see \citealt{Fuller_2022}).

Our stellar models exhibit a uniform initial rotation profile at various angular velocities. In \textsc{Mesa}, rotation is treated in a shellular approximation \citep{Zahn1992_Shellular_rotation, Meynet_Maeder_1997_shellular_rot}, implying that isobaric shells have uniform angular velocity.
As the models evolve, they lose mass and hence angular momentum. In addition, the angular momentum is redistributed internally, causing the rotation profile to deviate from that of the initial solid body. In \textsc{Mesa}, the transfer of angular momentum is modeled in a diffusive approximation. This is accomplished by contribution from ordinary diffusion, meridional circulations (e.g., \citealt{Eddington1925, Sweet1950}), and the magnetic field instability-induced Spruit-Tayler dynamo \citep{Spruit2002}. 
Although the mapping between the initial and final angular velocities is nonlinear, we choose a sufficient number of models (900+) such that it results in stars with different levels of final angular momentum reserves.

For wind mass loss, we follow the Dutch wind mass loss prescription as a combination of \cite{deJager_1988, Vink_2001} and \cite{Nugis_Lamers_2000} with the \texttt{Dutch\_scaling\_factor} = 0.25 and consider only a single metallicity with $Z = 10^{-4}$ with the solar scaled relative metal abundances adopted from \cite{Asplund:2009}.
Since the results in Section~\ref{sec: results} are calculated so that they remain only a function of the final mass and the angular momentum of the collapsing star, the metallicity of the star is not the variable of interest. However, metallicity would play a direct role in determining the amount of mass and angular momentum retained by the star and also its critical rotation rate \citep{Marchant:2023}. 
For certain exceptional cases (e.g., those where angular momentum is sourced from the tidal influence of a binary companion), the angular momentum content can remain large even at larger metallicity; therefore, our approach remains valid for such cases as well.
We also include rotation-induced mass loss as $ \dot{M}(\Omega)=\dot{M}(0) (1- \Omega / \Omega_{\text {crit }})^{- \xi}$ with $\xi = 0.43$ \citep{Langer1998}. As the angular velocity $\Omega$ approaches its critical value $\Omega_{\rm crit}$,  this expression begins to diverge. Hence, we set the variable \texttt{implicit\_mdot\_boost} = 0.1 in \textsc{Mesa} to artificially boost the mass loss rate until the rotation falls below critical.

% \vspace{-12pt}
\subsection{Formalism for collapsar dynamics} \label{sec: formalism}

Our goal is to determine the final mass and spin of the black hole resulting from the collapse of a rotating star of a given mass and angular momentum distribution. For such a purpose, the detailed nature of the collapse and the dynamics of the evolution of the accretion disk might not be necessary. As such, we focus on the effective (i.e., time-averaged) dynamics of the accretion flow, as detailed below. Unless explicitly stated, we use geometric units with $G, c = 1$ throughout this section.

% \vspace{-10pt}
\subsubsection{Background geometry}

As an initial condition, we assume that a black hole of some mass promptly forms during the initial collapse of the stellar core. The spacetime region of this black hole near its horizon (i.e., ignoring the effect of the remaining matter, which will be approximately be distributed in a spherically symmetric fashion) can be modeled using the Kerr metric ($g_{ab}$) whose line element in the Boyer-Lindquist coordinates $(x^a :=~t,r, \theta, \phi)$ reads
% # ------- This is the full metric -----
\begin{equation}
    \begin{aligned}
    ds^2 & = g_{ab}dx^a dx^b = -\left( 1 - \frac{2Mr}{\Tilde{\rho}^2} \right) dt^2 - \frac{4 Mar \sin^2{\theta}}{\Tilde{\rho}^2}d\phi dt \\
    & + \frac{\Tilde{\rho}^2}{\Delta} dr^2 +  \Tilde{\rho}^2 d\theta^2 + \left( r^2 + a^2 + \frac{2M r a^2 \sin^2 \theta}{\Tilde{\rho}^2} \right) \sin^2 \theta d\phi^2 \,,
    \end{aligned}
\label{eq: full Kerr metric}
\end{equation}
where $a = J/M; \; \Delta = r^2-2Mr+a^2$ and $\Tilde{\rho}^2 = r^2+a^2\cos^2\theta$. Here, $M, J$ are the black hole's mass and angular momentum, respectively.
As the remainder of the stellar matter falls inwards, the metric evolves due to a change in the black hole mass and spin. 

For simplicity, we assume that at any given time, there is a negligible amount of matter in the equatorial plane such that the effect of the disk mass on the form of Eq.~\ref{eq: full Kerr metric} can be ignored. Additionally, we assume that the outer shells have a nearly spherical configuration such that the spacetime inside the innermost infalling shell can be assumed to be flat (except for the effect due to the black hole).
There are two Killing vector fields $(\xi, \eta)$ associated with this stationary, axisymmetric spacetime namely $\xi = (1,0,0,0)\partial_t$ and $\eta = (0,0,0,1)\partial_{\phi}$. They correspond to the conservation of energy and the axial component of the angular momentum, respectively.

% \vspace{-7pt}
\subsubsection{The rotating fluid}

The final state of a massive star evolved through \textsc{Mesa} constitutes the initial state of the following analysis. Prior to collapse, we assume (in accordance with the \textsc{Mesa} model) that the stellar fluid comprises of matter, with only the axial component of angular momentum being nonzero.
Let $u^a$ be the 4-velocity of a fluid element of this star with components ($u^t, u^r, u^\theta, u^\phi$) in Boyer-Lindquist coordinates\footnote{The Latin indices run from 0 to 3 and the Greek indices from 1 to 3.}. When the core experiences a collapse, the outer shells will initially be in a free fall. During this phase, we model stellar matter as a fluid with the stress-energy tensor. 
\begin{equation}
    T^{ab} = (\rho + \epsilon + P) u^a u^b + Pg^{ab} \,,
    \label{eq: stress-energy tensor}
\end{equation}
where $g^{ab}$ (or equivalently $g^{-1}$) is the ``inverse'' metric, $P$ is the pressure and $\rho, \epsilon$ are the rest mass and internal energy density. We further assume that $\rho \gg (\epsilon + P)$. This reduces Eq.~\ref{eq: stress-energy tensor} to a dust equation of state. 
Since the pressure approaches the energy density only in the inner region of the disk, our approximation remains valid while the matter freely falls to form the disk.

\vspace{-7pt}
\subsubsection{Initial formation of the disk} \label{sec: formation of the disk}

Just prior to infall, the shell is located at a large $r$ value with $u^a = (\gamma, 0,0,\gamma \Omega)^a \partial_a$  such that $\Vert u^t \Vert \gg  \Vert u^\mu \Vert$ and therefore the shell falls from a near-rest configuration. Here, $\gamma = dt/d \lambda$, $\lambda$ being the fluid element's proper time.
For simplicity and compatibility with the \textsc{Mesa} models, we assume that the stellar matter collapses in a shellular fashion with angular velocity $\Omega = d\phi/dt$ (measured at $r\rightarrow \infty$).  The mass contained within such a shell at time $t$ can be written as
\begin{equation}
    dM = \int_{0, 0}^{2\pi, \pi} T^{ab} u_a n_b dV \,,
    \label{eq: dM}
\end{equation}
where $dV = \sqrt{g_{\Sigma_t}} dr d\theta d\phi$ and the integral is performed over $\theta, \phi$ only. Here $g_{\Sigma_t} = {\rm det}(g_{\mu \nu})$, $g_{\mu \nu}$ being the induced metric on the spacelike hypersurface $\Sigma_t$ at some time $t$ calculated by setting $dt = 0$ in the full Kerr metric $g_{ab}$ (e.g., \citealt{paschalidis2017rotating}). Additionally, $n_b =  (- dt / \sqrt{- g^{-1} (dt, dt)}, 0, 0, 0) = (- dt/\sqrt{- g^{tt}}, 0, 0, 0)$ is the unit 1-form normal to $\Sigma_t$ (where the 1-form $dt$ is the dual basis associated with the timelike Killing vector field $\xi = \partial_t$). The negative sign in front of $dt$ is chosen to keep $n_b$ future-directed.
% 
% Under the assumption that $\Vert u^t \Vert \gg  \Vert u^\mu \Vert$,  we may approximate the resulting energy-momentum density current as
% % 
% \begin{equation}
%     T^{ab}u_a \approx - \frac{{\left(a^{2} \cos^2 {\theta} - 2 \, M r + r^{2}\right)} \gamma^{2}}{a^{2} \cos^2 {\theta} + r^{2}} \rho u^t \,,
% \end{equation}
% % 
% 
 Thus, under the assumption that $\Vert u^t \Vert \gg  \Vert u^\mu \Vert$, Eq.~\ref{eq: dM} yields 
% 
% \begin{equation}
% \begin{aligned}
%    dM & = \int_{0, 0}^{2\pi, \pi} \rho \frac{\left(a^{2} \cos^{2} \theta+r^{2}\right)^{\frac{3}{2}} \sin \theta}{\sqrt{a^{2} \cos^{2} \theta -2 m r+r^{2}}} dr d\theta d\phi  \\
%    & \approx \int_{0, 0}^{2\pi, \pi} \gamma \rho r^2 \sin \theta  \, dr d\theta d\phi \,, %\quad r \gg M \,.
%     \label{eq: mass fallback}
%    \end{aligned}
% \end{equation}

\begin{equation}
\begin{aligned}
   dM & = \int_{0, 0}^{2\pi, \pi}  {(a^{2} \gamma^{3} \cos^2 {\theta}  - 2 \, \gamma^{3} M r + \gamma^{3} r^{2})} \rho \sin{\theta} \, dr d\theta d \phi \,. \\
    \label{eq: mass fallback}
\end{aligned}
\end{equation}
For $r \gg M$, one may approximate the above integral as
\begin{equation}
    dM = \int_{0, 0}^{2\pi, \pi} \gamma^3 \rho r^2 \sin \theta  \, dr d\theta d\phi \,,
    \label{eq: mass in shell}
\end{equation}
and for $\gamma = 1$, we recover the nonrelativistic variant of $dM$.
The angular momentum contained in the shell with mass $dM$ can be written as
\begin{equation}
\begin{aligned}
    dJ & = \int_{0, 0}^{2\pi, \pi} T^{ab} \eta_a n_b dV  \\
    & = \int_{0, 0}^{2\pi, \pi}  \gamma^{2} \rho  (a^{4} \Omega \cos^2{\theta} + 2a^{2} M \Omega r \sin^2{\theta} + a^{2} \Omega r^{2} \cos^2{\theta} \\
    & \quad \quad  \quad \quad   + a^{2} \Omega r^{2} + \Omega r^{4} - 2 a M r) \sin^3{\theta} \, dr d\theta d \phi \,, \\
    % & = \frac{8}{15} \pi  \gamma^{2} \rho \left(a^{4} \Omega + 6a^{2} \Omega r^{2} + 5 \Omega r^{4} + 8a^{2} Mr \Omega - 10a Mr \right) dr
    \label{eq: angular momentum fallback}
\end{aligned}
\end{equation}
which, for $r \gg M$ can be approximated as 
\begin{equation}
    dJ = \int_{0, 0}^{2\pi, \pi} \gamma^2 \rho \Omega r^4 \sin^3{\theta} \, dr d\theta d \phi \,.
\end{equation}
Again, for $\gamma = 1$, we recover the nonrelativistic variant of $dJ$ (cf. \citealt{Batta_2019}). Next, the condition $g_{ab} u^a u^b = -1$ can be used to estimate the corresponding $\gamma$ in Eq.~\ref{eq: mass fallback} and \ref{eq: angular momentum fallback} resulting in two solutions for $\gamma$ with the positive one being
\begin{equation}
    \gamma = \left(  \frac{r^2 a^2  +  \cos^2{\theta}}{r^2 - 2Mr + a^2 \cos^2{\theta}}   \right)^{1/2} \,.
\end{equation}
In our numerical setup, we set $\theta = \pi /4$ in the above equation, owing to the 1D nature of our approach.

Prior to disk formation, we assume that all mass and its associated angular momentum freely falls into the black hole, respecting the condition $a < M$ (see Section~\ref{sec: equilib spin} later). We are interested in finding the stellar radius $r_0$ at which the infalling shell has sufficient angular momentum to first form an accretion disk. This occurs when the (averaged) specific angular momentum $\ell$ of the collapsing shell in the equatorial plane equals the specific angular momentum of the innermost stable circular orbit (ISCO) $\ell_{\rm isco}$ of the Kerr metric.
At this point, the matter first circularises at the ISCO before plunging into the black hole.
Using Eq.~\ref{eq: mass fallback} and \ref{eq: angular momentum fallback}, we can calculate the specific angular momentum in the mass shell as
\begin{equation}
    \ell(r) = dJ/dM \,.
    \label{eq: specific AM}
\end{equation}
Additionally, the angular momentum per unit rest mass at radius $r$ of a circular orbit in the equatorial plane of the Kerr metric assuming prograde orbit is \citep{bardeen1972rotating}
\begin{equation}
    u_\phi (r) = \frac{ M^{1 / 2}\left(r^2 - 2 a M^{1 / 2} r^{1 / 2}+a^2\right)}{r^{3 / 4}\left(r^{3 / 2}-3 M r^{1 / 2} + 2 a M^{1 / 2}\right)^{1 / 2}} \,.
    \label{eq: specific AM in Kerr geometry}
\end{equation}
Choosing $r = r_{\rm isco}$ in Eq.~\ref{eq: specific AM in Kerr geometry}, the condition
\begin{equation}
    \ell(r_0) = u_{\phi} (r_{\rm isco})
    \label{eq: condition for disk formation}
\end{equation}
then gives us the stellar radius $r_0$ at which the infalling shell has sufficient angular momentum to first form an accretion disk. Since Eq.~\ref{eq: specific AM in Kerr geometry} is defined per unit rest mass, we accordingly change Eq.~\ref{eq: specific AM} to bring it to a similar form to assist in the comparison in Eq.~\ref{eq: condition for disk formation} above. 
The coordinate time interval $\Delta t~=t(r_0) - t(r_{\rm isco})$ meanwhile gives us the disk formation time post-core-collapse (within an error margin of a sound crossing time as the information of the collapse would take this much time to reach $r_0$). As we discuss in Appendix~\ref{sec: free_fall time}, $\Delta t$ can be well approximated by the corresponding Newtonian free fall time. Therefore, to simplify our calculations, we use the latter approach for calculating $\Delta t$.

\vspace{-7pt}
\subsubsection{Evolution of the disk} \label{sec: evolution of the disk}

Once Eq.~\ref{eq: condition for disk formation} is satisfied, the subsequent infalling matter would first be assimilated into the disk before being accreted by the black hole or blown away in the wind. For the accretion phase, we restrict ourselves to the equatorial plane of the black hole, where the accretion disk will reside. 
The line element in the equatorial plane $(\theta = \pi/2)$  reads
\begin{equation}
    d s^2 = -\left(1 - \frac{2M}{r} \right) d t^2 - \frac{4aM}{r} dt d\phi  + \frac{r^2}{\Delta} d r^2 + \frac{A}{r^2} d\phi^2 \, ,
\label{eq: Kerr metric in plane}
\end{equation}
% 
% where $a = J/M; \; \Delta \equiv r^2-2 M r+a^2; \; A \equiv r^4+r^2 a^2+2 M r a^2$.
% 
where $A \equiv r^4+r^2 a^2+2 M r a^2$.
At a given time $t$, let the disk have mass $M_d$ and angular momentum $J_d$. The corresponding effective disk radius $r_d$ can be calculated by equating
\begin{equation}
    u_\phi (r_d) = J_d/M_d \,,
\end{equation} 
where the term on the LHS is given in Eq.~\ref{eq: specific AM in Kerr geometry}.
The mass and angular momentum evolution of the disk can be written as
\begin{equation}
    \dot{M}_d = \dot{M}_{\rm fb} - \dot{M}_{\rm lost}; \quad \dot{J}_d = \dot{J}_{\rm fb} - \dot{J}_{\rm lost}  \,,
\end{equation}
where the dot represents derivative w.r.t. $t$. Here $\dot{M}_{\rm fb}, \dot{J}_{\rm fb}$ represent the mass and the angular momentum assimilation rate into the disk due to fallback and can be calculated using Eq.~\ref{eq: mass fallback} and \ref{eq: angular momentum fallback}. 
Meanwhile, the mass and angular momentum dissipation rate from the disk ($\dot{M}_{\rm lost}, \dot{J}_{\rm lost}$) will be calculated below.
% % 
% To calculate the $\gamma$ term in Eq.~\ref{eq: mass fallback} and \ref{eq: angular momentum fallback},

The infalling matter in the disk is accreted on a viscous timescale. During this process, to account for the effect of wind outflow, we model the local accretion rate as a power law \citep{Blandford_1999, Kumar_2008}
\begin{equation}
    \frac{ dM_{\rm acc} (r)}{d \tau}  = \frac{M_d}{\tau_{\rm vis}} \left( \frac{r}{r_d}\right)^s ;  \quad (r_t < r < r_d) \,,
    \label{eq: local accretion rate}
\end{equation}
where $s \in [0.3,0.8]$ is a free parameter (e.g., \citealt{Kumar_2008, Yuan_Narayan_2014}) which we assume to be $s = 0.5$ (e.g., \citealt{Kumar_2008}). We interpret Eq.~\ref{eq: local accretion rate} to model the accretion rate as measured from a locally nonrotating frame of reference ($r, \theta = $ const.). Such observers would have zero angular momentum (apart from the effect of frame dragging) and are also known as zero angular momentum observers (ZAMO). Here, $\tau_{\rm vis}$ is the viscous timescale (as measured by ZAMO) at the characteristic radius $r_d$ 
and is modeled using the \cite{Shakura_Sunyaev_1973} $\alpha$-viscosity prescription as \citep{abramowicz1996advection, gammie1998advection}
\begin{subequations}
    \begin{equation}
    \tau_{\rm vis} \approx  {2} / { (\alpha \ \Tilde{\Omega}) } ; \quad  {\Tilde{\Omega}} = \Omega_K - \omega \,,
    \end{equation}
    \vspace{-0.3cm}
    \begin{equation}
    \Omega_K = \frac{u^\phi}{u^t} = \frac{M^{1/2}}{r_d^{3/2} + aM^{1/2}}; \quad \omega = - \frac{g_{t \phi}}{g_{\phi \phi}} \,.
    \label{eq: Omega_K}
    \end{equation}
\end{subequations}
Above $\Omega_K, \omega$ is the angular velocity of the fluid element (in a circular orbit) and the angular velocity of the frame dragging, respectively, as measured by an observer at radial infinity. In addition, $\Tilde{\Omega}$ is the angular velocity of the fluid as measured by a ZAMO \citep{bardeen1972rotating}. It is expected that $\alpha \in [0.01, 0.1]$ with theory supporting the lower value and observation the larger one \citep{King_2007}. Here, we take $\alpha = 0.01$. A variation in the value of $\alpha$ would influence the maximal value of $a^* := a/M$ \citep{Sadowski_2011} and therefore may affect the evolution of the black hole (see Section~\ref{sec: equilib spin} and \ref{subsec: Variation of the fiducial parameters}).
To transform Eq.~\ref{eq: local accretion rate} into the Boyer-Lindquist frame, we convert the ZAMO proper time to the coordinate time as (e.g., see section 10.7 in \citealt{gourgoulhon2018geometry})
\begin{equation}
    d \tau = \frac{dt}{\sqrt{- g^{tt}}} ; \quad 1/\sqrt{- g^{tt}} = \sqrt{\Delta} \left(r^2 + a^2 + \frac{2a^2Mr}{\Tilde{\rho}^2}   \right)^{-1/2} \,.
\end{equation}

In Eq.~\ref{eq: local accretion rate}, for $r \leq r_t$ - where $r_t$ is some transitional radius below which the disk undergoes efficient neutrino cooling, resulting in no wind mass loss and hence efficient accretion - we set $s = 0$. We calculate $r_t$ by setting the RHS of Eq.~\ref{eq: local accretion rate} to ${ 10^{-2.5} r} / {(2M)} \, M_\odot {\rm s}^{-1}$ (e.g., \citealt{Kohri_2005, Chen_Beloborodov_2007, Kumar_2008}).
When $r > r_t$, the disk enters the advection-dominated accretion flow regime (ADAF, \citealt{Narayan_Yi_1994, Narayan_McClintock:2008_review}), where most of the disk mass is blown away in winds. The net mass loss rate from the disk - owing to outflows and black hole accretion - yields
\begin{equation}
    \dot{M}_{\rm lost} = \dot{M}_{\rm acc}(r < r_t) + \underbrace{ \int_{r_t}^{r_d} d \dot{M}_{\rm acc}}_{\dot{M}_{\rm wind}} \,,
\end{equation}
where $\dot{M}_{\rm wind}$ can be evaluated as 
\begin{equation}
    \dot{M}_{\rm wind} = \frac{M_d}{\tau_{\rm vis}} \frac{s}{r_d^s} \int_{r_t}^{r_d} \frac{r^{s-1}}{\sqrt{ -g^{tt}}}  dr \,.
\end{equation}
We resort to numerical integration to further evaluate the above equation. Meanwhile, the net angular momentum loss rate yields
\begin{equation}
    \dot{J}_{\rm lost} =  \dot{J}_{\rm BH} + \underbrace{\int_{r_t}^{r_d} u_\phi(r) d \dot{M}_{\rm acc}(r)}_{\dot{J}_{\rm wind}} \,,
    \label{eq: J_wind unintegrated}
\end{equation}
where $\dot{J}_{\rm BH}$ in defined Eq.~\ref{eq: angular momentum accretion by BH} and  $\dot{J}_{\rm wind}$, represents the angular momentum lost in the wind outflow, assuming that the latter removes the corresponding specific angular momentum from its location of origin. This integral can be evaluated as
\begin{equation}
\begin{aligned}
    \dot{J}_{\rm wind} & =  \frac{M_d}{\tau_{\rm vis}} \frac{s}{r_d^s} \int_{r_t}^{r_d} \frac{u_\phi(r)}{\sqrt{ -g^{tt}}} r^{s-1} dr \\
    & \approx \frac{M_d}{\tau_{\rm vis}} \frac{s}{r_d^s} \left[   \frac{2 a M r^{s-1}}{1-s}+\frac{2 \sqrt{M} r^{s+\frac{1}{2}}}{2 s+1}  \right]_{r_t}^{r_d} \,.
    \label{eq: J_winds}
\end{aligned}
\end{equation}
To calculate the last term, we dropped terms $\mathcal{O}(a/r^2)$ within the integral. This is more accurate than assuming a Schwarzschild geometry with the dropped terms becoming increasingly less unimportant as $a$ decreases or $r$ increases. For the case where $a = 0$, Eq.~\ref{eq: J_winds} reduces to the one given in \cite{Kumar_2008}. Moreover, for the current work, we find Eq.~\ref{eq: J_winds} to also provide a fair match to the numerically integrated value of the $\dot{J}_{\rm wind}$ term in Eq.~\ref{eq: J_wind unintegrated}.

Finally, to incorporate the influence of the anti-parallel relativistic jets on the accretion flow, we assume that at disk formation, the polar region making a half-opening angle of 
\begin{equation}
    \theta_c =  \Tilde{\eta}^{3/4} \left(\frac{L_j}{3 r_*^2 \rho_* c^3}\right)^{1 / 4} \,
\end{equation}
with the rotational axis evolves into a relativistic cocoon and is promptly removed from the star. Here $L_j, r_*, \rho_*$ is the jets' luminosity, star's radius, and its mean density, respectively. Additionally, $\Tilde{\eta}$ is the ratio of the speed of light to the jets' head speed and we assume $\Tilde{\eta} = 10/3$, e.g. \citealt{Nakar:Piran:2017} - see Appendix~\ref{sec: estimating theta_c} for more information on estimating $\theta_c$.

% A typical assumed value for the relativistic cocoon half opening angle is 0.5 radian $\approx 28.65^\circ$, e.g., \cite{Ramirez-Ruiz:2002, Nakar:Piran:2017}. 

% \vspace{-8pt}
\subsection{Evolution of black hole with a Novikov-Thorne disk}

In the early phase, while the disk is yet to form, we assume that the mass freely falls into the black hole without losing energy or angular momentum.
After an initial disk formation, the subsequent disk dynamics is modeled as discussed in Section~\ref{sec: evolution of the disk}. This is analogous to modeling the accretion flow as consisting of a relativistic, radiatively efficient, geometrically thin accretion disk that lacks the presence of large-scale magnetic field interactions with the black hole, also known as \cite{Novikov_Thorne_1973} disk (NTD).

For such an accretion flow, once the mass reaches $r_{\rm isco}$, it is assumed to be directly accreted by the black hole, adding mass at a rate
\begin{equation}
    \dot{M}_{\rm BH} =  u_t (r_{\rm isco}) \dot{M}_{\rm acc}(r_t) \,,
    \label{eq: mass accretion rate}
\end{equation}
where (also note that $\dot{M}_{\rm acc}$ remains unchanged for $r \leq r_t$)
\begin{equation}
     u_t (r) =\frac{r^{3 / 2}-2 M r^{1 / 2} + a M^{1 / 2}}{r^{3 / 4}\left(r^{3 / 2}-3 M r^{1 / 2} + 2 a M^{1 / 2}\right)^{1 / 2}} \,,
    \label{eq: specific energy}
\end{equation}
is the energy  per unit rest mass of a circular orbit at the radius $r = r_{\rm isco}$ \citep{bardeen1972rotating}.
This also adds angular momentum to the black hole at a rate\footnote{Note that we have added the subscript ``BH'' to the variables to explicitly associate them to the black hole.}
\begin{equation}
    \dot{J}_{\rm BH} = u_{\phi} (r_{\rm isco}) \dot{M}_{\rm acc} (r_t) \,,
    \label{eq: angular momentum accretion by BH}
\end{equation}
where, following \cite{bardeen1972rotating}, $r_{\rm isco}$ is 
\begin{equation}
\begin{aligned}
    & r_{\rm{isco}} = M \left[3+z_2 -  \sqrt{((3-z_1)(3+z_1+2 z_2)}\right] \,, \\
    & z_1=1+ (1-a^2 / M^2)^{1 / 3}((1+a/M)^{1 / 3}+(1-a/M)^{1 / 3}) \,,\\
    & z_{2} = \sqrt{3a^2 /M^2 + z_{1}^{2}} \,.
\end{aligned}
\end{equation}

\vspace{-7pt}
\subsubsection{Equilibrium spin of the black hole} \label{sec: equilib spin}

In principle, as the black hole accretes mass (and hence angular momentum), its spin parameter will gradually approach the limit $a  = M$. However, on considering the decelerating impact of disk-emitted photons on the black hole, \cite{Thorne_1974} found that the dimensionless spin parameter of the black hole $a^*:= a/M$ should converge to a value of $a^* = 0.9978$.
Since both Eq.~\ref{eq: specific AM} and \ref{eq: specific energy} are functions of $a$, hence, the equilibrium value of $a^*$ becomes important\footnote{E.g., for $a^* = 0.9978, 0.9994$, $u_t(r_{\rm isco}) \approx 0.68, 0.58$ respectively.}.
More recently, \cite{Sadowski_2011} recalculated this limit for the case of (advection dominated, optically and geometrical thick) slim accretion disks and found $a^* =  0.9994$ for $\alpha = 0.01$ and a $10 \, \times$ super-Eddington mass accretion rate. Since these conditions also manifest in collapsar physics, in this work, we use the latter equilibrium value for $a^*$ when modeling the accretion flow using an NTD. This means that we cap the maximum black hole spin to $a^* = 0.9994$ in our simulation.

% unless the inner region of the disk is penetrated by a strong magnetic field in which case the black hole spins down to nearly zero, as we discuss next.

\vspace{-7pt}
\subsection{Evolution of black hole with a magnetically arrested disk} \label{sec: MAD}

Magnetic fields are considered to be crucial in extracting the rotational energy of the Kerr black hole to power the anti-parallel relativistic jets. Of particular interest are the predictions of general relativistic magnetohydrodynamic (GRMHD) simulations of a magnetically arrested disk (MAD) state (e.g., \citealt{Tchekhovskoy_2011}). Here, the accretion disk is threaded by a strong vertical magnetic field that is dragged inward by the accretion flow, resulting in an accumulation of magnetic flux near the black hole horizon. These magnetic fields can play a central role in extracting the rotational energy of the black hole \citep{Penrose_1971, Blandford_Znajek_1977}.

To calculate the effect of MAD on the black hole, in conjunction with Section~\ref{sec: evolution of the disk}, here we follow the approach of \cite{Moderski_Sikora_1996}, which accounts for the energy and angular momentum loss of the black hole due to the Blandford-Znajek mechanism as
\begin{subequations}
    \begin{equation}
    \dot{J}_{\rm BH} = {\ell_{\rm in} \dot{M}_{\rm acc, in}} -  \frac{P_{\mathrm{EM}}}{k \Omega_{\rm H}} \,;
    \end{equation}
    % \vspace{-0.5cm}
    \begin{equation}
        \dot{M}_{\rm BH} = u_{t, \mathrm{in}}\dot{M}_{\rm acc, in}  - \frac{P_{\mathrm{EM}}}{c^{2}} \,,
    \end{equation}
    \label{eq: MAD variables}
\end{subequations}
\hspace{-2pt}where $\Omega_{\rm H} = r_{\rm H} / (M_{\rm BH}a)$ is the angular frequency of the black hole's event horizon (measured at $r \rightarrow \infty$) and $r_{\rm H} = M + \sqrt{M^2 - a^2}$. For the MAD state, the variables in Eq.~\ref{eq: MAD variables} need to be determined from GRMHD simulations, for which we use the recent analytical fits provided in the 3D GRMHD study of \cite{Lowell_2023} as
\begin{subequations}
    \begin{equation}
       P_{\rm EM} = \eta_{\rm EM} \dot{M}_{\rm acc, in } c^2; \quad \eta_{\rm EM} = 1.063{a^*}^4 + 0.395{a^*}^2 \,,
    \end{equation}
    \vspace{-0.5cm}
    \begin{equation}
       \ell_{\rm in} = 0.86 GM/c; \quad \quad \quad u_{t, \rm in} = 0.97 \,,
    \end{equation}
    \vspace{-0.5cm}
    \begin{equation}
        k = \Omega_{\rm F}/ \Omega_{\rm H} = \rm{min}(0.1 + 0.5a^*, 0.35) \,.
    \end{equation}
    \label{eq: MAD parameters}
\end{subequations}
Here, $P_{\rm EM}$ is the electromagnetic flux leaving the black hole, which is primarily contained in the jets. Also, $\Omega_{\rm F}$ is the angular frequency of the magnetic field lines at the horizon. The subscript ``in'' corresponds to the inner end of the accretion disk in the simulations of \cite{Lowell_2023}, which is not necessarily at the ISCO.

\vspace{-7pt}
\subsection{Electromagnetic energy of explosion} \label{sec: gravitational redshift}

The energy radiated during a GRB may arise either due to the viscous dynamics of the accretion flow, e.g., via the \cite{Blandford_Payne:1982} mechanism (as is the case for an NTD) or a combination of the former plus the Blandford-Znajek mechanism, the latter robbing the black hole of its energy and angular momentum (e.g., for the case of a MAD). This energy would have to climb the gravitational potential well of the black hole in order to be observed by a distant observer.

% --------------------------------------------------------
\begin{figure}
    \centering
    \vspace{-5pt}
    \includegraphics[width = 1\linewidth]{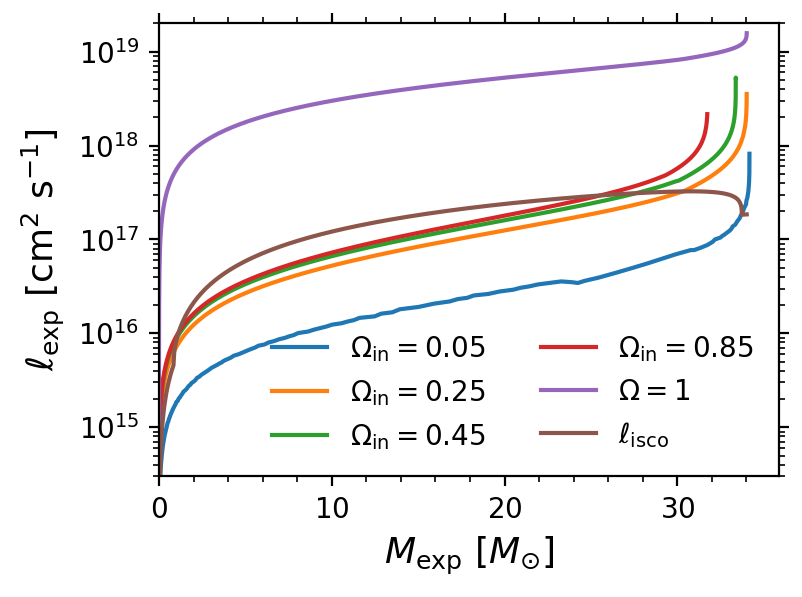}
    \vspace{-18pt}
    \caption{The angular momentum profile of a helium star evolved from the onset of core-helium burning with an initial mass of $35M_\odot$ and a range of initial angular velocities (normalized by the critical angular velocity) $\Omega_{\rm in}$. The purple curve with $\Omega = 1$ represents the profile of a star that is rotating at its critical/Keplerian rate (at the point of explosion) and has been plotted for comparison with other models. We note that this does not consider the effect of radiation pressure on the stellar layers and thus overestimates the true value of $\Omega = 1$.  $\ell_{\rm isco}$ represents the specific angular momentum at the ISCO if the \textit{entire} inner mass collapses to form a black hole (respecting $a^* \leq 0.9994$). Note that models with large $\Omega_{\rm in}$ lose a much larger amount of angular momentum in stellar winds such that their $\ell_{\rm exp}$ converges to a value that is more than an order of magnitude smaller than that of $\Omega = 1$.}
    \label{fig: angular_momentum_profile}
\end{figure}
% -----------------------------------------------------------

\vspace{-7pt}
\subsubsection{Gravitational redshift of the emitted energy}
We account for the energy lost due to the gravitational redshift of the radiation when measured by an observer at radial infinity as follows. 
For a stationary observer at radius $r$, its 4-velocity is $u^a = \sqrt{-1/g_{tt}} \cdot \xi^a$. The specific energy of a radially moving photon as measured by this observer at $r$ is 
\begin{equation}
    E(r)  = - g_{ab} u^a u^b = - \sqrt{-1/g_{tt}} g_{ab} u^a \xi^b \,.
\end{equation}
Now,  $g_{ab} u^a \xi^b$ should remain conserved along the photon's geodesic, given $\xi$ is Killing. Thus, the observed specific energy at $r \rightarrow \infty$ is given by $E_\infty = - g_{ab} u^a \xi^b$.  Consequently,  at a given time, the specific energy of radiation at infinity is related to its value at some radius $r$ as
\begin{equation}
    E_{\infty}  = E(r) \sqrt{-g_{tt}} \,,
    \label{eq: E_inf}
\end{equation}
where $g_{tt}$ is calculated at the radius $r$ and $E(r) = 1 - u_t(r)$ is the total specific energy emitted by the fluid element during its decent to radius $r$ from infinity. For computational ease, we evaluate the $g_{tt}$ term in Eq.~\ref{eq: E_inf} assuming $r= 2r_{\rm acc}$, where $r_{\rm acc}$ is the radius at which the element is accreted (e.g., $r_{\rm acc} = r_{\rm isco}$ for an NTD). This is a fair approximation as roughly half of the energy would be emitted on each side of this value.
Meanwhile, if the energy is sourced from the spin of the black hole (e.g., in the MAD state), one may set $r = r_{\rm ergo}$, where $r_{\rm ergo} = M + \sqrt{M^2 - a^2 \cos^2{\theta}}$ is the outer radius of the ergosphere of the black hole, and assume $\theta = \pi/2$. 
However, under the assumption that the photons move radially outward from the ergosphere (as we did in Eq.~\ref{eq: E_inf}), the latter becomes a surface of infinite redshift.  Since actual photons would contain angular momentum and hence be able to escape the ergosphere to infinity; thus, to avoid the issue of infinite redshift, for our current purpose, we somewhat ad hocly set $r=1.01 \times r_{\rm ergo}$ in Eq.~\ref{eq: E_inf} when the energy is sourced from the spin of the black hole.

% ---------------------------------------------------
\begin{figure}
    \centering
    \vspace{-10pt}
    \includegraphics[width = 1\linewidth]{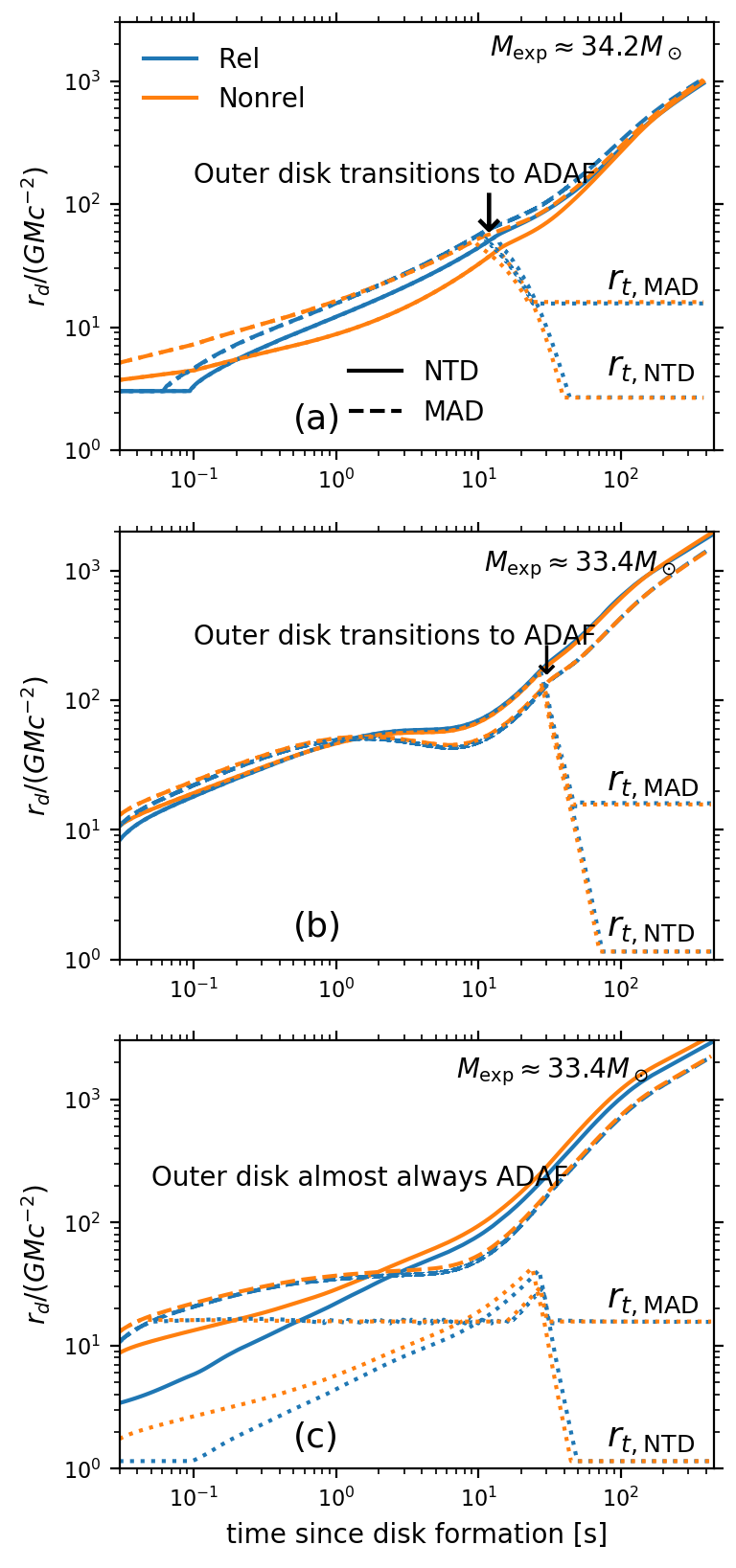}
    \vspace{-18pt}
    \caption{The evolution of $r_d$ post disk formation. The blue (orange) curve represents a relativistic (non-relativistic) model. The solid (dashed) curve represents the NTD (MAD) state. The dotted line represents the transitional radius $r_t$ for \textit{both} the models. Time axis shows the coordinate (proper) time when a relativistic (non-relativistic) approach is employed. Fig. (a) shows the $r_d$ evolution of a star with $M_{\rm exp} = 34.2M_\odot$ and $\Omega_{\rm in} = 0.25$, and Figs. (b) and (c) consider $M_{\rm exp} = 33.4M_\odot$ and $\Omega_{\rm in} = 0.45$. Both stars have same initial mass of $35M_\odot$ (see Fig.~\ref{fig: angular_momentum_profile}). Since the blue and orange curves do not visually differ in Fig. (b), in Fig. (c), we artificially move $r_t$ inward by a factor of 5 from its actual location to demonstrate the effect of Kerr geometry on the accretion flow.}
    \label{fig: r_d evolution}
\end{figure}
% --------------------------------------------------------

\vspace{-10pt}
\subsubsection{Isotropic equivalent energy of explosion}

We classify the collapsar explosion as a potential source of GRB if the total isotropic equivalent energy $E_{\rm iso}$ received at radial infinity is greater than $10^{51}$ erg \citep{Perley_2016}. This implies
\begin{subequations}
    \begin{equation}
    E_{\rm iso, NTD} = \frac{\beta}{f} \int_{t_{\rm exp}} E_\infty \, \dot M_{\rm acc}(r_{\rm isco}) \, dt  > 10^{51} \, \rm{erg} \,,
\end{equation}
\begin{equation}
\begin{aligned}
     E_{\rm iso, MAD} = \frac{1}{f} \int_{t_{\rm exp}} & \left[ \beta E_\infty \, \dot M_{\rm acc, in}  \quad + \right. \\  & \left. \sqrt{-g_{tt}} \eta_{\rm EM} \dot{M}_{\rm acc, in } c^2 \right] dt  > 10^{51} \, \rm{erg} \,,
\end{aligned}
\end{equation}
\end{subequations}
where the integral is performed over the total time interval of the explosion, and the subscript of $E_{\rm iso}$ represents the state of the accretion flow. Additionally, $\beta = 0.01$ is the fraction of energy that is released in the form of radiation (with the rest lost in neutrinos). In addition, we set $\beta = 1$ when the radiation energy is sourced from the spin of the black hole. The term $f:= \hat{\Omega} / 4\pi$ represents the fractional probability that the radiation (mainly emitted in the polar jets) aligns with the observer's line of sight, where $\hat{\Omega}$ is the total solid angle spanned by the radiation and can be calculated as
% 
% ------ Deriving the above equation --------
\begin{equation}
\begin{aligned}
     \hat{\Omega} & = 2 \int_0^{2\pi} \int_0^\theta \sin{\theta} d\theta d\phi \\
     & = 2\int_0^{2\pi} (-\cos{\theta} + 1) d\phi = 4\pi(1 - \cos{\theta}) \,,
\end{aligned}
\end{equation}
where $\theta$ is radiation's half-opening angle w.r.t. the black hole's rotational axis. Following the calibration of \cite{Bavera_2022}, we take $f = 0.05$.

\vspace{-5pt}
\section{Result} \label{sec: results}

By numerically implementing the formalism discussed in Section~\ref{sec: numerical methods}, here, we assess the impact of the progenitor star's mass and angular momentum reservoir on the properties of the subsequently produced black hole, such as its mass and spin. For the latter, we provide fitting functions that can be readily used by future studies (in particular population synthesis) where the pre-collapse nature of the progenitor star is known. But before that, we first present a brief comparison of the output of the current 1D relativistic implementation to the non-relativistic one.

% --------------------------------------------------------
\begin{figure*}
    \vspace{-10pt}
    \begin{multicols}{2}
     \centering
    \includegraphics[width = .95\linewidth]{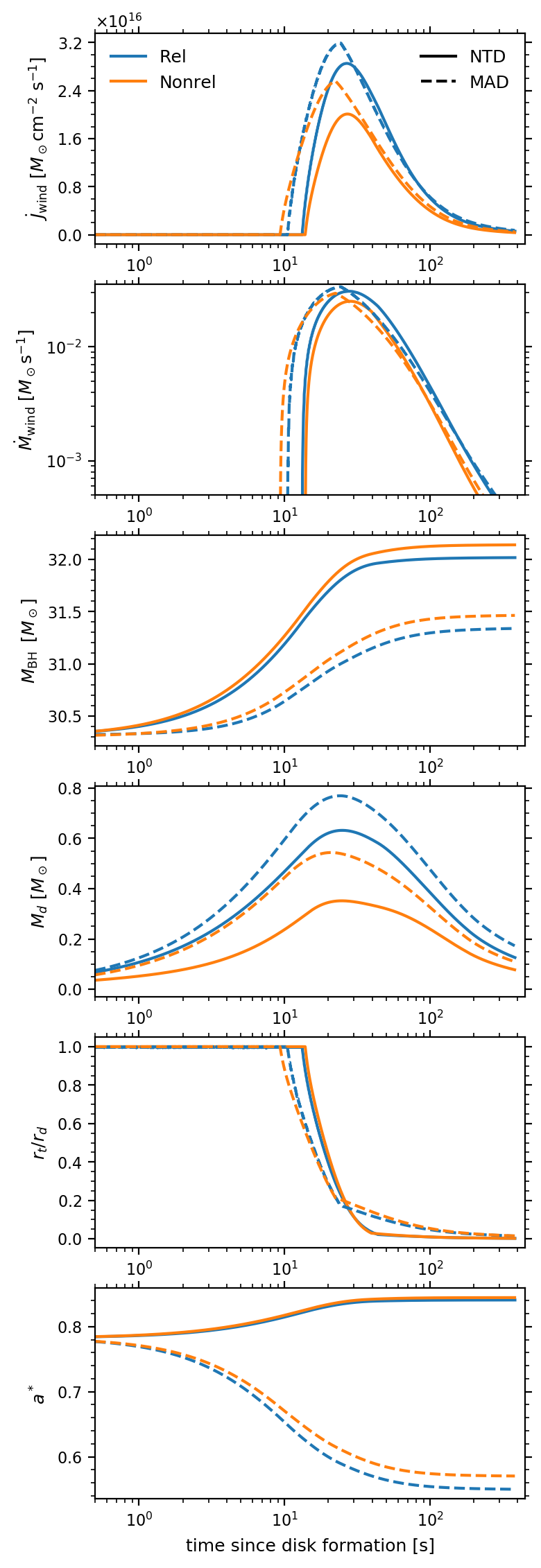} \par
    \includegraphics[width = .95\linewidth]{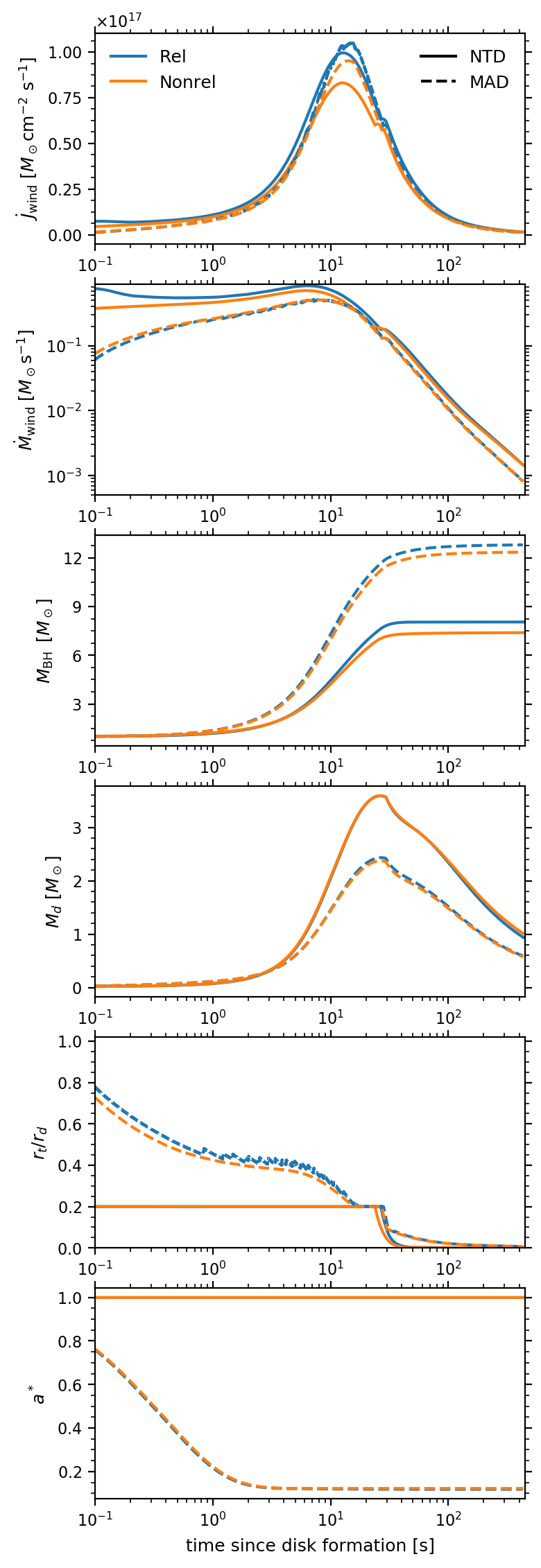} \par
    \end{multicols}
    
    \vspace{-15pt}
    \caption{Time evolution of the relevant parameters since the onset of disk formation. LHS figure shows the model with $M_{\rm exp} \approx 34.2M_\odot$ and $\Omega_{\rm in} = 0.25$. The RHS figure is the same as LHS but for $M_{\rm exp} = 33.4 M_\odot$ and $\Omega_{\rm in} = 0.45$. The legend has the same meaning as in Fig.~\ref{fig: r_d evolution}. Here $M_{\rm BH}, a^*$  represent the mass and spin of the black hole, $\dot{M}_{\rm wind}, \dot{J}_{\rm wind}$ the rate of loss of mass and angular momentum in disk winds, $M_d$ the disk's mass, $r_t$ the transitional radius below which no wind mass loss occurs, and $r_d$ the characteristic disk radius. Similar to Fig.~\ref{fig: r_d evolution}, for the RHS figure, we have artificially shifted $r_t$ inward by a factor of 5 (for both relativistic and non-relativistic approaches). Note the change in the range of the x-axis for the RHS figures.}
    \label{fig: disk evolution}
\end{figure*}
% --------------------------------------------------------

\vspace{-10pt}
\subsection{Comparison to the non-relativistic approach} \label{subsec: rel vs non-rel}

 A non-relativistic variant of the NTD formalism described in Section~\ref{sec: numerical methods} is presented in \cite{Kumar_2008} and then numerically implemented in \cite{Fuller_2022}. 
 To perform a suitable comparison with the above-discussed relativistic treatment, we also modify the non-relativistic formalism to match the requirement set here for the NTD (see Section~\ref{sec: equilib spin}) and the MAD state (see Section~\ref{sec: MAD}). 
 To compare the two approaches, we choose a representative stellar model with an initial mass $35M_\odot$ that rotates at a range of initial angular velocities $\Omega_{\rm in}$. The corresponding angular momentum profile of these models at the moment of collapse is illustrated in Fig.~\ref{fig: angular_momentum_profile}.
 
As shown in Fig.~\ref{fig: r_d evolution}, during the initial stage of collapse, when near-horizon physics is important, the two treatments (i.e., relativistic and nonrelativistic) may visibly differ from each other. We find this difference to be evident if the black hole has a lower value of $a^*$ at the onset of disk formation (e.g., Fig.~\ref{fig: r_d evolution}a with the spin evolution later shown in Fig.~\ref{fig: disk evolution}). On the other hand, for $a^* \approx 1$ (Fig.~\ref{fig: r_d evolution}b), both treatments lead to similar predictions for the location of $r_t$ and $r_d$. This is because a progenitor that can produce a black hole with $a^* \approx 1$ at the onset of disk formation contains a considerable amount of angular momentum in its infalling mass shells. 
This causes $r_d$ to assume a large value and thus be located at a relatively far location from the black hole (e.g., see Fig.~\ref{fig: r_d evolution}b). 
% 
% This causes the disk to form at a relatively far location from the black hole (e.g., see Fig.~\ref{fig: r_d evolution}b). 
% 
For such scenarios, the difference between the relativistic and non-relativistic treatments only becomes apparent if we artificially decrease the value of the transitional radius $r_t$ (and thus also the value of $r_d$ as $r_t$ influences the latter), bringing it closer to the horizon.
This is demonstrated in Fig.~\ref{fig: r_d evolution}c where on decreasing $r_t$ from its actual location by a factor of 5 (for both the treatments), both $r_t$ and $r_d$ begin to differ substantially early on when the disk is near the horizon in the two treatments for the NTD. 
On the other hand, during later stages, the disk circularizes at a relatively larger distance from the horizon, thus diminishing the effect of the Kerr geometry. Meanwhile, for the MAD state, the inner disk radius is set by Eq.~\ref{eq: MAD parameters}b, which (for the current model) result in a relatively large inner disk radius. Since $r_t$ cannot be made smaller than the inner disk radius, in the MAD state, the lowest possible value of $r_t$ in Fig.~\ref{fig: r_d evolution}c occurs when $r_t = r_{\rm in}$ for both the (relativistic and non-relativistic) treatments. As $r_{\rm in}$ lies relatively far away from the black hole thus, the disk properties for the relativistic and nonrelativistic treatments of the MAD state do not show much difference.

The evolution of the most relevant variables associated with the model discussed in Fig.~\ref{fig: r_d evolution} is presented in Fig.~\ref{fig: disk evolution}. We note that the time axis represents the coordinate (proper) time when a relativistic (non-relativistic) approach is used. Thus, for the variables that involve a time derivative, a meaningful comparison between the two treatments is best achieved at a relatively large distance from the black hole (e.g., see Fig.~\ref{fig: infall coordinate time}). For the current purpose, this is the case for most part of the collapse. 
% 
% 
% While the LHS plot in Fig.~\ref{fig: disk evolution} shows the true disk evolution, the RHS plot again has its $r_t$ value changed.
One can check that the tiny but noticeable difference in the {black hole 
properties} under the relativistic and non-relativistic approaches in Fig.~\ref{fig: disk evolution} can mainly be attributed to the relative difference in the location of $r_t$ for the two treatments. Meanwhile, the relative amplification in the value of disk mass $M_d$, and the rate of mass and angular momentum loss in winds $\dot{M}_{\rm wind}$ and $\dot{J}_{\rm wind}$ (for the relativistic model) can mainly be attributed to the presence of the $\gamma$ term in Eq.~\ref{eq: mass fallback} and \ref{eq: angular momentum fallback}.

% Finally, we note that including the effect of gravitational redshift - as discussed in Section~\ref{sec: gravitational redshift} - reduces the magnitude of $E_{\rm iso}$, as measured by an observer at radial infinity by $79 \pm 2 \%$, $85 \pm 5\%$ for an NTD and MAD state, respectively.

% --------------------------------------------------------------
\begin{figure*}
    \centering
    \vspace{-7pt}
    \includegraphics[width = 0.95\linewidth]{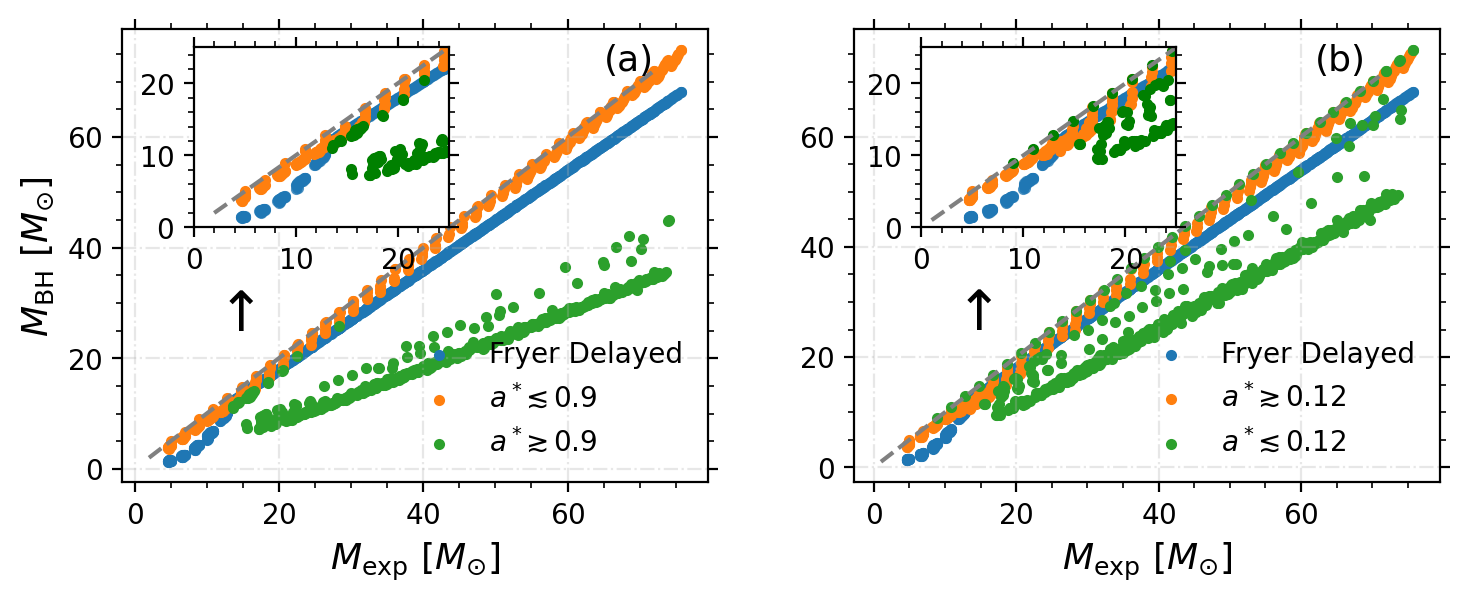}
    \vspace{-10pt}
    \caption{The masses of black holes as a function of the final mass of their progenitor star (evolved till core carbon depletion). The blue points show the black hole masses resulting from the \textit{Fryer Delayed} remnant mass function. The label $a^*$ represents the dimensionless spin of the black hole. Figure (a) is for the NTD state, while Figure (b) is for the MAD state. In both cases (depending on the angular momentum reservoir of the progenitor star), for $M_{\rm exp} \gtrsim 15 M_\odot$, the black hole mass function experiences a bifurcation into two separate branches. Note that the resulting green branches in Figures (a) and (b) show opposite trends in their values of $a^*$.  The green points (i.e., those with $a^* \lesssim 0.12$ ) on the upper branch in Figure (b) are those systems that had low angular momentum already at formation and not because of a GRB-induced spin down.}
    \label{fig: M_exp_vs_M_BH}
\end{figure*}
\begin{figure*}
    \centering
    \vspace{-5pt}
    \includegraphics[width = 0.95\linewidth]{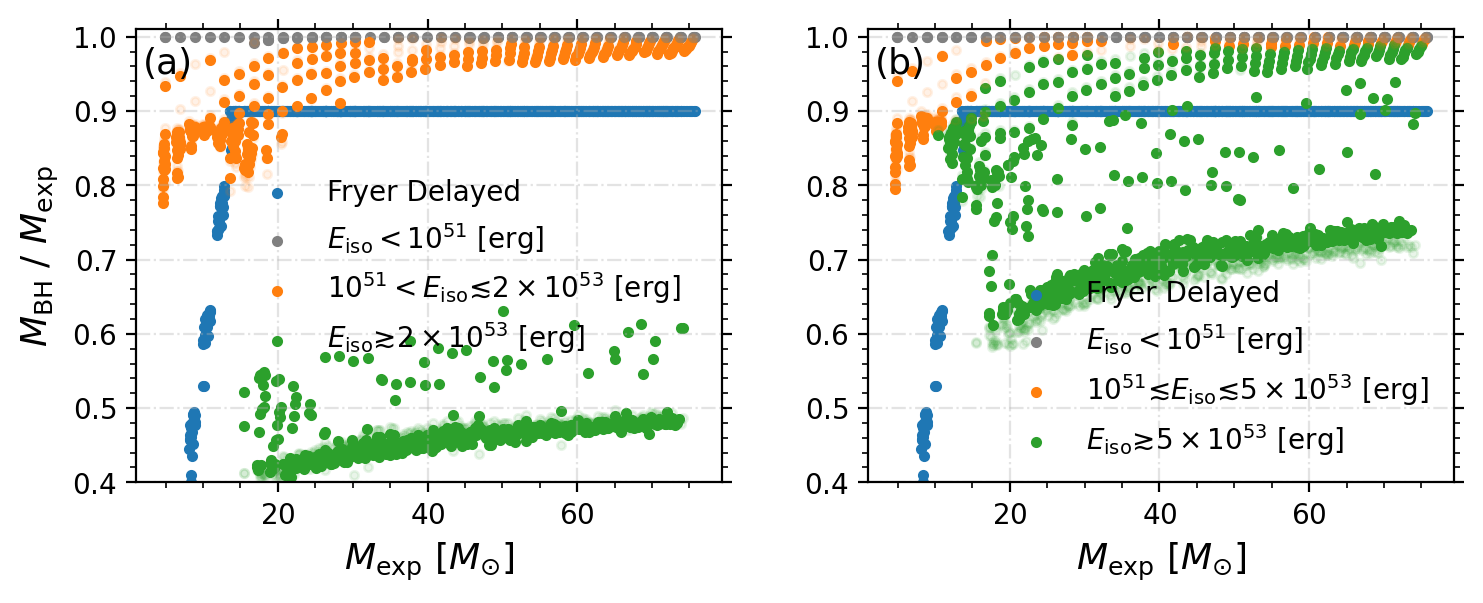}
    \vspace{-10pt}
    \caption{Same as Fig.~\ref{fig: M_exp_vs_M_BH} but now the y-axis shows the ratio of the black hole's mass to the pre-explosion mass of its progenitor star and the legend shows the isotropic-equivalent energy $E_{\rm iso}$ of the GRB. Additionally, for comparison, we have overplotted the corresponding $M_{\rm BH}/M_{\rm exp}$ values resulting from the non-relativistic approach in lighter color data points. These are not visible in the LHS figure but can be seen in the RHS one. The gray points are those systems that we do not classify as a potential source of GRB.}
    \label{fig: Mexp_vs_MBH_ratio}
\end{figure*}
% --------------------------------------------------------------

\vspace{-10pt}
\subsection{Masses of black holes born from rotating progenitors} \label{sec: BH mass spectrum}

In this section, we explore the impact of the progenitor star's mass and angular momentum reservoir on the mass of the subsequently formed black hole.

\subsubsection{Masses of black holes as a function of $M_{\rm exp}$}

To this end, Fig.~\ref{fig: M_exp_vs_M_BH} and \ref{fig: Mexp_vs_MBH_ratio} show the final masses of black holes as a function of the pre-explosion mass $M_{\rm exp}$ of their progenitor \textsc{Mesa} models. For comparison, we use the \textit{Fryer Delayed} remnant mass function as a source of reference. Additionally, we also overplot the corresponding black hole masses resulting from the non-relativistic approach in Fig.~\ref{fig: Mexp_vs_MBH_ratio}.

Fig.~\ref{fig: M_exp_vs_M_BH} and \ref{fig: Mexp_vs_MBH_ratio} shows that there is a threshold value of $a^*$ and $E_{\rm iso}$ respectively beyond which the remnant mass trend bifurcates into two separate branches. The magnitude of such an $a^*$ or $E_{\rm iso}$ and the amount of bifurcation in the resulting black hole masses is a function of the hydrodynamical nature of the accretion flow. In the case of an NTD, the resulting drop in the mass of black holes can at times be larger than $50\%$, see Fig.~\ref{fig: Mexp_vs_MBH_ratio}a. Meanwhile, the MAD state in Fig.~\ref{fig: Mexp_vs_MBH_ratio}b tends to produce black holes with relatively larger masses compared to the Novikov-Thorne flow. {This can be attributed to the difference in the location of $r_t$ for the two accretion flows. For example, Fig.~\ref{fig: r_d evolution}a and \ref{fig: r_d evolution}b show that once the outer disk transitions to an ADAF (as also discussed in Section~\ref{subsec: rel vs non-rel}), the location of $r_t$ for a MAD state is relatively further from the horizon (compared to the NTD state). Thus, in the latter case, the black hole accretes more mass, resulting in larger final masses. 
Nevertheless, the mass of the black holes is still lower than those resulting from the convection-enhanced neutrino heating explosion, as showcased by the \textit{Fryer delayed} mechanism. 

For stars with relatively smaller reservoirs of angular momentum, the collapse is semi-direct, without much dependence on the nature of the accretion flow. For such stars, we find the masses of the black holes to be typically larger than those predicted by the \textit{Fryer delayed} mechanism. Although for such slow rotators, one might expect convection-enhanced neutrino heating to additionally contribute towards explosion energetics. However, the scarcity of observations of supernova explosions of stars with mass $\gtrsim 18 \, M_{\odot}$ suggests that direct collapse might be possible \citep{Smartt:2015, Adams:2017}. Numerical studies (e.g., \citealt{Murguia_Berthier:2020}) also align with observations, suggesting that massive stars might experience a direct collapse without an ensuing electromagnetic counterpart.

We also observe that, unlike the case of NTD, for the MAD state, the threshold $E_{\rm iso}$ does not guarantee a downward shift in the black hole masses. For example, in Fig.~\ref{fig: Mexp_vs_MBH_ratio}b, we find very luminous GRBs with $E_{\rm iso} \gtrsim 5 \times 10^{53}$ erg that also produce relatively massive black holes. 
Such explosions result from progenitor stars that only develop a black hole accretion disk during the later stage of their collapse (e.g., the scattered green points in Fig.~\ref{fig: time of disk formation}), hence minimizing their chances of losing substantial mass in ADAF winds. Nonetheless, the black hole is still able to produce energetic GRB jets as the latter efficiently feeds on its spin angular momentum. Meanwhile, the black hole does not significantly spin down. To achieve the latter, a near maximally rotating black hole needs to accrete $\approx 20\%$ of its initial mass (e.g., \citealt{Jacquemin_Ide:2023}), which is not the case for the aforementioned models. 
A demonstration of this process can also be seen in the bottom LHS of Fig.~\ref{fig: disk evolution}, which shows the spin evolution of a $34.4M_\odot$ black hole post-disk-formation. Here, the black hole's spin is not significantly depleted as the disk only forms near the end of the accretion phase.

\begin{figure*}
    \centering
    \vspace{-5pt}
    \includegraphics[width = 0.95\linewidth]{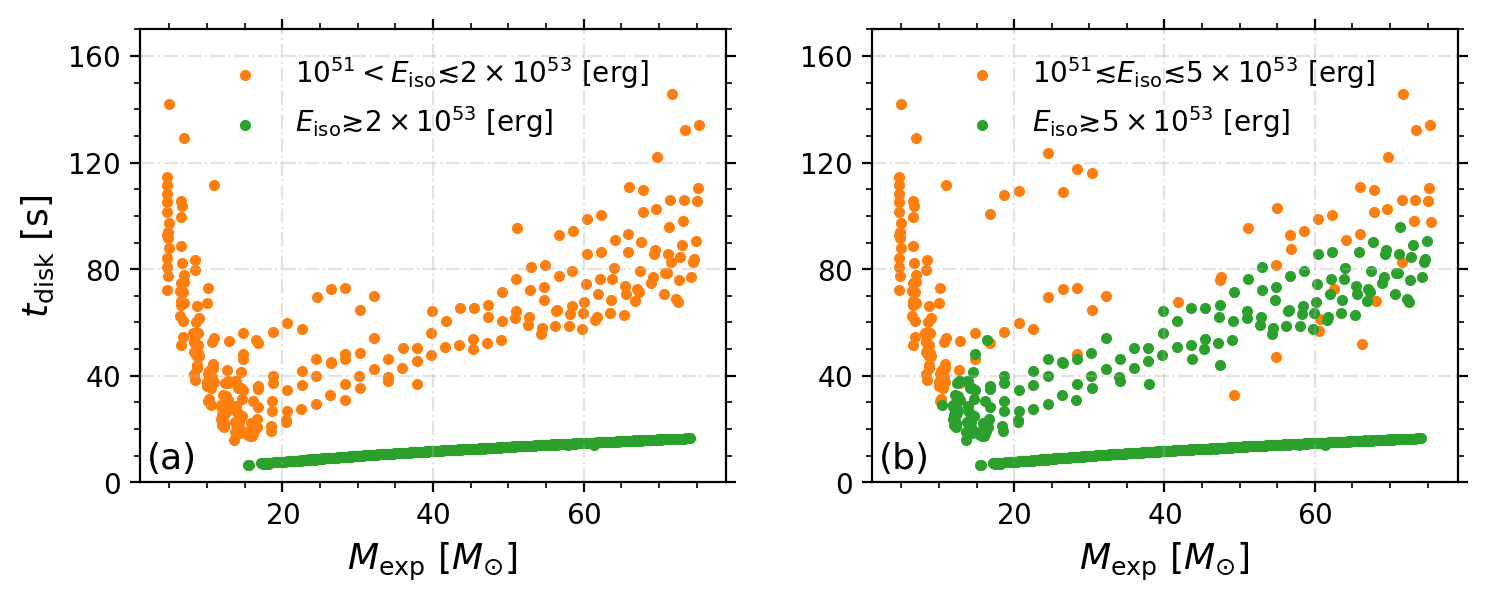}
    \vspace{-5pt}
    \caption{The disk formation time for the various models. The LHS figure is for the NTD state while RHS is for the MAD state. The legend is similar to Fig.~\ref{fig: Mexp_vs_MBH_ratio}. Note that the \textit{scattered} green points (with relatively larger $t_{\rm disk}$ values) on the RHS figure are the same \textit{scattered} models plotted in green color in Fig.~\ref{fig: Mexp_vs_MBH_ratio}b} 
    \label{fig: time of disk formation}
\end{figure*}

% --------------------------------------------------------------------
\begin{figure*}
    \centering
    \vspace{-5pt}
    \includegraphics[width = 1\linewidth]{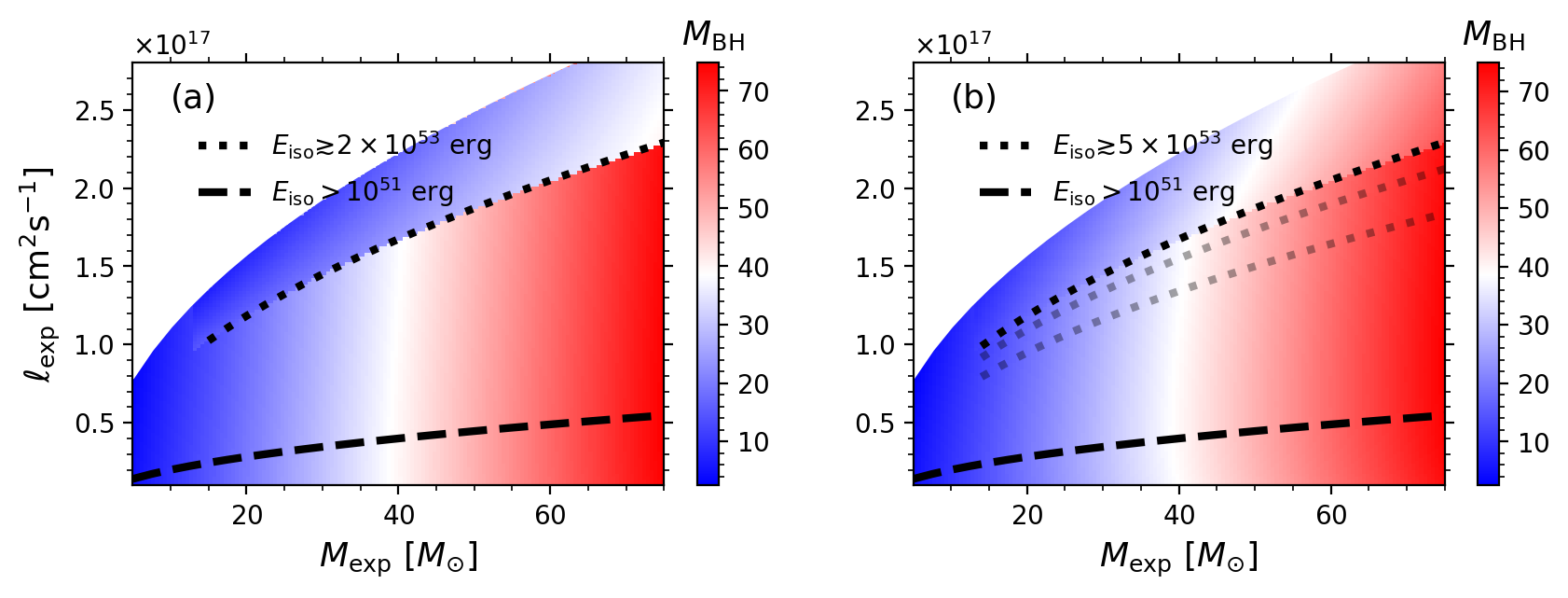}
    \vspace{-10pt}
    \caption{The masses of black holes as a function of their progenitor star's mass and angular momentum reservoir. The figures are drawn from the fits provided in Eq.~\ref{eq: 2d_MBH_fit}. The x-axis represents the mass of the star and the y-axis its specific angular momentum at the point of core carbon depletion. Figure (a) is for the NTD state, while Figure (b) is for the MAD state. The region in gray dotted lines in Figure (b) corresponds to the very luminous systems in Fig.~\ref{fig: Mexp_vs_MBH_ratio}(b) that have $E_{\rm iso} \gtrsim 3 \times 10^{53}$ erg but do not satisfy Eq.~\ref{eq: threshold AM} to experience significant ADAF winds.} 
\label{fig: 2D_hist_MBH}.
\end{figure*}
% --------------------------------------------------------------------

\subsubsection{Masses of black holes as a function of $M_{\rm exp}$ and $\ell_{\rm exp}$}

To better quantify the dependence of the black hole properties on the progenitor's structure, we now consider a broader parameter space constituting the mass of the star prior to explosion $M_{\rm exp}$ and its (averaged) specific angular momentum $\ell_{\rm exp}$.

To this end, as shown in Fig.~\ref{fig: 2D_hist_MBH}, we find an approximate relation between $M_{\rm exp}$ and $\ell_{\rm exp}$ (i.e., the black dashed line)
\begin{equation}
    \bar{\ell}_{\rm exp} \gtrsim (4 \times 10^{3}M_{\rm exp})^{1/2}; \quad \text{ for \,\,}  M_{\rm exp} \gtrsim 5 M_{\odot} \,,
    \label{eq: minimum AM}
\end{equation}
where we first expect a progenitor star to experience a GRB. Similarly, the relation
\begin{equation} 
    \bar{\ell}_{\rm exp} \gtrsim (7 \times 10^{4}M_{\rm exp})^{1/2};  \quad \text{ for \,\,} M_{\rm exp} \gtrsim 15 M_{\odot}
    \label{eq: threshold AM}
\end{equation}
which is depicted by the black dotted lines in the figure, marks the onset of the region where we first expect to see a strong deviation in the black hole masses $M_{\rm BH}$. In other words, Eq.~\ref{eq: threshold AM} separates the parameter space belonging to the two branches shown earlier in Fig.~\ref{fig: M_exp_vs_M_BH} and \ref{fig: Mexp_vs_MBH_ratio}. 

The parameter space lying above the boundary set by Eq.~\ref{eq: threshold AM} is the region where the black hole mass experiences a sudden downward shift in trend. 
This shift can be attributed to the large reservoir of angular momentum of the progenitor, which causes the accretion disk to form soon after the onset of the collapse. After sufficient time, the accretion flow of such a collapse transitions into the ADAF regime, resulting in a strong mass loss. Moreover, NTD-hosting black holes in Fig.~\ref{fig: 2D_hist_MBH}a also lose a considerable amount of mass in neutrinos from the inner region (owing to the Urca processes), hence efficiently carrying away energy with them. 

Finally, the region plotted in gray dotted lines in Fig.~\ref{fig: 2D_hist_MBH}b corresponds to the previously discussed very luminous systems in Fig.~\ref{fig: Mexp_vs_MBH_ratio}b that have $E_{\rm iso} \gtrsim 5 \times 10^{53}$ erg but do not satisfy Eq.~\ref{eq: threshold AM}. These systems have lower angular momentum reservoirs; however, since, for the MAD state, the GRB jets are primarily powered by the black hole spin, a strong jet is still produced.
We also note that above a certain value of $M_{\rm exp}$ pair-instability supernova might occur ( \citealt{Fowler_and_Hoyle1964, Rakavy_Shaviv:1967}, see Section~\ref{sec: PISN} for more information), thus truncating the black hole masses function; however, we provide the remnant masses for an extended range of $M_{\rm exp}$ values and leave it on the reader to decide the location of the pair-instability cut-off. Similarly, the effect of pulsation pair-instability \citep{Woosley_PPISNe_2007} on the star's $M_{\rm exp}$ would need to be determined before using the current results.

In the following, we now provide the fit functions that have been used to plot Fig.~\ref{fig: 2D_hist_MBH}. A comparison between these fits and the true values is provided in Fig.~\ref{fig: error_MBH} and \ref{fig: MBH_error_comparison}.
For readability, let us first define ${\ell}_{\rm exp}:= \bar{\ell}_{\rm exp} \times 10^{14} \text{ [cm}^2{\rm s}^{-1}]$. 
Then, the corresponding function for the masses of the black holes (in $M_{\odot}$) used for generating Fig.~\ref{fig: 2D_hist_MBH} can be approximated within $8\%$ error (with $90\%$ credibility; see Fig.~\ref{fig: error_MBH}) as
% 

% \begin{widetext}
\begin{subequations}
\begin{equation}
    \begin{array}{l}
    M_{\rm BH}^{\rm NT}  = \left\{\begin{array}{ll} 
     0.333 \sqrt{M_{\rm exp}\bar{\ell}_{\rm exp}}  
     + 1.435 \sqrt{M_{\rm exp}} 
    - 1.99 \sqrt{\bar{\ell}_{\rm exp}} \\ 
    - 0.46 M_{\rm exp} - 0.0155 \bar{\ell}_{\rm exp}  + 55.2; \,\, \text{if Eq.~\ref{eq: threshold AM} holds}, \\ \\
      0.018 \sqrt{M_{\rm exp}\bar{\ell}_{\rm exp}}  
     - 0.07 \sqrt{M_{\rm exp}} 
    - 0.043 \sqrt{\bar{\ell}_{\rm exp}} \\ 
    + 0.99 M_{\rm exp} - 0.003 \bar{\ell}_{\rm exp} + 0.45; \,\,  \text{ otherwise}, 
    \end{array}\right.
    \end{array}
\end{equation}
\begin{equation}
    \begin{array}{l}
    M_{\rm BH}^{\rm MA} = \left\{\begin{array}{ll} 
     0.04 \sqrt{M_{\rm exp}\bar{\ell}_{\rm exp}}  
    - 0.517 \sqrt{M_{\rm exp}} 
    + 0.003 \sqrt{\bar{\ell}_{\rm exp}} \\
    + 0.98 M_{\rm exp}  
    - 0.006 \bar{\ell}_{\rm exp}  + 1.22; \,\, \text{ if Eq.~\ref{eq: threshold AM} holds}, \\ \\
    0.055 \sqrt{M_{\rm exp}\bar{\ell}_{\rm exp}}  + 1.08  \sqrt{M_{\rm exp}} 
    + 0.92 \sqrt{\bar{\ell}_{\rm exp}} \\ 
    + 0.834  M_{\rm exp} - 0.03 \bar{\ell}_{\rm exp}  - 8.51; \,\, \text{ otherwise}, 
    \end{array}\right. 
    \end{array}
\end{equation}
\label{eq: 2d_MBH_fit}
\end{subequations}
% \end{widetext}
% 
where the superscripts on $M_{\rm BH}$  represent the assumption on the nature of the accretion flow during the collapse, and the constant coefficients have appropriate units to make the LHS and RHS dimensionally consistent. 
% 

% --------------------------------------------------------------------
\begin{figure*}
    \centering
    \vspace{-5pt}
    \includegraphics[width = 1\linewidth]{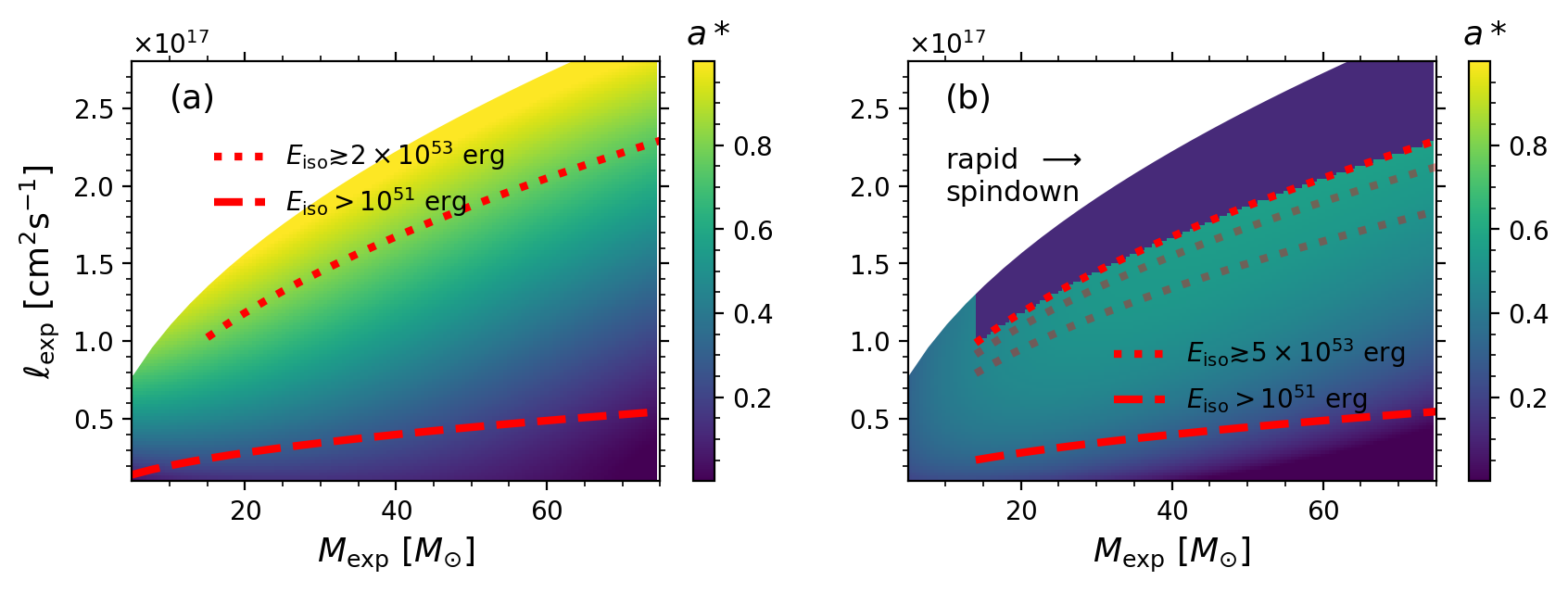}
    \vspace{-10pt}
    \caption{Same as Fig.~\ref{fig: 2D_hist_MBH} but now showing the dimensionless spin parameter $a^*$ of black holes. Figures are drawn using the fits provided in Eq.~\ref{eq: 2d_spin_fit}.}
    \label{fig: 2D_hist_spin}
\end{figure*}

% --------------------------------------------------------------------

% \vspace{-14pt}
\subsection{Spin of black holes born from rotating progenitors}

Similar to Fig~\ref{fig: 2D_hist_MBH}, Fig.~\ref{fig: 2D_hist_spin} shows the dependence of black hole spin on the progenitor's $M_{\rm exp}$ and $\ell_{\rm exp}$. Here, Fig.~\ref{fig: 2D_hist_spin}a shows that in the case of an NTD, for any given value of $M_{\rm exp}$ the black hole spin increases monotonically as a function of $\ell_{\rm exp}$. Moreover, beyond a certain threshold value of $\ell_{\rm exp}$ (i.e., the dotted red line in Fig.~\ref{fig: 2D_hist_spin}a, see Eq.~\ref{eq: threshold AM}), the black hole typically acquires a spin of $a^* \gtrsim 0.9$. This coincides with the region where the mass of the black hole experiences a sudden downward bifurcation in Fig.~\ref{fig: 2D_hist_MBH}a.

 While in the case of an NTD, the GRB energetics fully result from the accretion dynamics, for the MAD state, the spin angular momentum of the black hole plays a direct role in powering the jets. Consequently, we expect a dramatic variation in the final spin of the black holes depending on the nature of the accretion flow. This can be seen in Fig.~\ref{fig: 2D_hist_spin}b, where the maximum attainable spin value for the MAD state is $a^* \approx 0.59$. This is because, now, the most rapidly spinning black holes fuel strong GRBs right after their formation, which efficiently spin them down to an equilibrium spin of $a^* \approx 0.12$ (cf. \citealt{Gottlieb:2023, Jacquemin_Ide:2023}).

As was done previously, we now provide the fit functions that have been used to plot Fig.~~\ref{fig: 2D_hist_spin}, with a comparison between these fits and the true values provided in Fig.~\ref{fig: error_MBH} and \ref{fig: spin_error_comparison}.
The spin magnitude of the black hole shown in Fig.~\ref{fig: 2D_hist_spin} can be approximated within 11\% error (with 90\% credibility) as
\begin{subequations}
\begin{equation}
     \begin{array}{l}
    a_{\rm NT}^* =  \left\{\begin{array}{ll} 
      -0.002\sqrt{M_{\rm exp}\bar{\ell}_{\rm exp}}  +0.0019\sqrt{M_{\rm exp}}  
     +0.048 \sqrt{\bar{\ell}_{\rm exp}} \\
     +0.004 M_{\rm exp} 
     -0.0004 \bar{\ell}_{\rm exp}  
     +1.09 \times10^{-6}M_{\rm exp}\bar{\ell}_{\rm exp} \\  -0.043;  \,\, \text{ if Eq.~\ref{eq: threshold AM} holds}, \\ \\
     -0.002 \sqrt{M_{\rm exp}\bar{\ell}_{\rm exp}}  +0.068 \sqrt{M_{\rm exp}}  
     +0.025 \sqrt{\bar{\ell}_{\rm exp}} \\
      - 0.006 M_{\rm exp} 
      + 0.0003 \bar{\ell}_{\rm exp}  
      - 2.034 \times10^{-6}M_{\rm exp}\bar{\ell}_{\rm exp} \\ - 0.222 ;
      
      \,\,  \text{ otherwise},
    \end{array}\right.
    \end{array}
\end{equation}
\begin{equation}
    \begin{array}{l}
    a_{\rm MA}^* =  \left\{\begin{array}{ll} 
    0.12; \qquad  \text{ if Eq.~\ref{eq: threshold AM} holds}, \\ \\
     {\rm min}(0.00036 \sqrt{M_{\rm exp}\bar{\ell}_{\rm exp}}  +0.066 \sqrt{M_{\rm exp}}  
     +0.0165\sqrt{\bar{\ell}_{\rm exp}} \\
      -0.0088 M_{\rm exp} 
      -2.036 \times 10^{-5} \bar{\ell}_{\rm exp}  
      -6.54 \times 10^{-7} M_{\rm exp}\bar{\ell}_{\rm exp}  \\ - 0.14, \, 0.59); \,\, \text{ otherwise}.
    \end{array}\right. 
    \end{array}
\end{equation}
\label{eq: 2d_spin_fit}
\end{subequations}
where the superscripts on $a^*$  represent the assumption on the nature of the accretion flow during the collapse. Like Eq.~\ref{eq: 2d_MBH_fit}, the constant coefficients have appropriate units to make the LHS and RHS dimensionally consistent. To ensure that the fit values remain positive, one should bound the values of $a_{\rm NT}^*$ and $a_{\rm MA}^*$ from below by zero.

\subsection{Domain of validity}

Although it is challenging to measure the spin of individual components in a merging BBH, the mass-weighted effective spin parameter $\chi_{\rm eff}$ of the binary can be relatively well estimated from the observed gravitational wave signal \citep{Abbott_O3a_2021, Abbott_O3b_2021}.  For a BBH comprising masses $M_1$ and $M_2$ with spins $a^*_1$ and $a^*_2$ aligned with their orbital angular momentum,
\begin{equation}
    \chi_{\rm eff} = \frac{M_1 a^*_1 + M_2 a^*_2}{M_1 + M_2} \,.
\end{equation}
Interestingly, recent findings suggest that most stellar mass black holes should be born with a negligible spin \citep{Eggenberger_2019, Fuller_2019} with values around $a^* \approx 0.01$ \citep{Fuller_Ma_2019}. 
Therefore, binaries with $\chi_{\rm eff} \gtrsim 0.01$ might have a distinct formation history. The prevalent methods for generating spinning black holes in isolated binaries involve either (i) tidal spin-up of a progenitor star within a tight binary, with a fraction of the acquired angular momentum subsequently inherited by the ensuing black hole (e.g., \citealt{deMink_Mandel2016, Marchant2016, Qin_2018, Bavera_2020}), or (ii) sufficient mass (and hence angular momentum) accretion onto the progenitor star (e.g., \citealt{Cantiello2007, Ghodla_2022_CHE}) or the first-born black hole (e.g., \citealt{Zevin_Bavera_2022}).

\vspace{-5pt}
\subsubsection{Spin-up via tidal interaction} \label{sec: spin up via tidal interaction}

Tidal spin-up can occur while both stars are on their hydrogen main-sequence due to the occurrence of chemically homogeneous evolution (CHE e.g., \citealt{deMink_Mandel2016, Marchant2016}) or when the primary star first evolves into a black hole with now having a tidally locked secondary star that is on its helium main-sequence with negligible hydrogen envelope (e.g., \citealt{Qin_2018, Bavera_2020, Fuller_2022}). 
Assuming that the companion star's spin period promptly synchronizes with its orbital period $P$, the value of such a star's (averaged) specific angular momentum $\ell$ can be estimated as
\begin{equation}
    \ell = \frac{I \omega}{M_{\rm star}} = \frac{2\pi (r_g r_*)^2}{P} \,; \quad I = M_{\rm star} (r_g r_*)^2,
    \label{eq: spin period}
\end{equation}
where $I, \omega, M_{\rm star}$ are the moment of inertia, orbital angular velocity and the mass of the star, respectively. Additionally, $r_*$ represents the radius and $r_g r_*$ its radius of gyration.

Using Eq.~\ref{eq: spin period} one can thus estimate the domain of validity of the fit functions provided in Eq.~\ref{eq: 2d_MBH_fit} and \ref{eq: 2d_spin_fit}.
For example, using the helium star models presented in Section~\ref{sec: MESA models}, the blue curve in Fig.~\ref{fig: domain_of_relavance} shows the threshold maximum orbital period of the binary at the onset of helium main-sequence required to sufficiently spin-up the star for it to satisfy Eq.~\ref{eq: threshold AM} by the time the star reaches core carbon depletion. Note that the metallicity of such stars needs to be low to minimize mass and, hence, angular momentum loss in stellar winds while the star evolves to core collapse (our helium star models have $Z = 10^{-4}$).  The orange curve, on the other hand, considers tidal synchronization till core helium depletion and, therefore, may be applicable to high metallicity stars as well. Post core helium depletion, the star with even larger metallicities will undergo relatively less mass loss and, hence, spin angular momentum loss and thus a strong tidal locking might not be necessary.

\vspace{-5pt}
\subsubsection{Spin-up via mass accretion}

Stellar spin-up via mass accretion might occur while the star is on its hydrogen main-sequence \citep{Cantiello2007, Ghodla_2022_CHE}. For such systems, we expect the stars to undergo a GRB if they satisfy Eq.~\ref{eq: minimum AM} at core carbon depletion. Additionally, the properties of the subsequently formed black hole could be highly affected if such systems also satisfy Eq.~\ref{eq: threshold AM} (see Fig.~\ref{fig: 2D_hist_MBH} and \ref{fig: 2D_hist_spin}). The latter is likely to be only valid in rapidly rotating low metallicity stars, i.e., $Z \lesssim \mathcal{O}(5 \times 10^{-4})$ \citep{Ghodla_2022_CHE}. However, in contrast to black holes born in tight binaries, black holes born from stars that undergo accretion-induced spin-up might not merge efficiently unless the supernovae kick results in compact and/or significantly eccentric orbits. 
This is because to efficiently spin up a star; one likely requires the formation of an accretion disk around the mass gainer (e.g., \citealt{Ghodla_2022_CHE}). However, for the formation of an accretion disk around the mass gainer, one needs to allow for a relatively larger separation between the binary components, which subsequently results in longer gravitational wave merger time. As such, although produced in relatively large numbers (e.g., \citealt{Ghodla_2022_CHE}), such black holes might seldom feature in the BBH mergers detected on Earth by current-generation detectors.

\vspace{-5pt}
\subsection{Variation of the fiducial parameters} \label{subsec: Variation of the fiducial parameters}

There are three main free parameters in the present work, - namely the power law index $s$ in Eq.~\ref{eq: local accretion rate}, the parameter $\Tilde{\eta}$ that controls the cocoon half opening angle $\theta_c$ (see Section~\ref{sec: evolution of the disk}), and the disk viscosity parameter $\alpha$ -  that can impact the black hole properties. While our choices of the fiducial values for these parameters (i.e., $s = 0.5, \Tilde{\eta} = 10/3$, and $\alpha = 0.01$) are motivated by previous studies, here we discuss the impact on the black hole mass and spin when these values are changed.

In addition to the fiducial values, we consider two variants for $s$, namely $s = 0.3, 0.8$, two variants for $\Tilde{\eta}$, namely $\Tilde{\eta} = 10/4, 10/2$ and one variant for $\alpha$, namely $\alpha = 0.1$. When $\alpha = 0.1$, we set the maximum spin of the black hole (Section~\ref{sec: equilib spin}) to $a^* = 0.9846$ \citep{Sadowski_2011} which then influences the location of the corresponding $r_{\rm isco}$.
In each attempt, we only change one variable while keeping the other two set to their fiducial values. However, we note that $\theta_c$ depends not only on the magnitude of $\Tilde{\eta}$ but also on that of $L_j$ at the time of disk formation. The latter can change during the variation of any of the aforementioned parameters, implying that the value of $\theta_c$ could also change.

The results achieved on variation of these parameters are presented in Appendix~\ref{sec: variation of fiducial parameters}. Fig.~\ref{fig: spt3} and \ref{fig: spt8} show that a variation in $s$ does not have a significant effect on the masses of the black hole compared to the fiducial case with the black hole masses (especially those corresponding to large $E_{\rm iso}$) being slightly smaller for a smaller value of $s$  and vice-versa.  Meanwhile, a variation in $\Tilde{\eta}$ does have a significant impact on the magnitude $M_{\rm BH}$. The corresponding results are presented in Fig.~\ref{fig: eta_2pt5} and \ref{fig: eta_5}. As one might expect, lowering the value of $\Tilde{\eta}$ results in less massive black holes as the jet now remains within the star for a longer period of time, hence efficiently exchanging energy with the stellar matter and displacing a larger volume from the polar region. 
Finally, Fig.~\ref{fig: alpha_pt1} shows that raising the value of the viscosity parameter to $\alpha = 0 .1$ only impacts the masses of the black holes when the accretion flow is in an NTD state. The inner radius of the accretion disk dictates the amount of mass that would be liberated in energy. Since for $\alpha = 0.1$, the maximum value of $a^* = 0.9846$ (in contrast to 0.9994 for $\alpha = 0.01$) thus, in the case of NTD, less energy is radiation away. Meanwhile, as discussed in Section~\ref{subsec: rel vs non-rel}, for the MAD state, the disk may truncate further away from ISCO. Hence, for the latter, changing $\alpha$ does not strongly influence the location of the inner disk radii.

Similarly, Fig.~\ref{fig: spin_spt3} and \ref{fig: spin_spt8} show the effect of $s$ variation, Fig.~\ref{fig: spin_eta_2pt5} and \ref{fig: spin_eta_5} show the effect of $\Tilde{\eta}$ variation and Fig.~\ref{fig: spin_alpha_pt1} show the effect of $\alpha$ variation on the black hole spin respectively. For all cases, the deviation in the black hole spin from the fiducial case is minimal. In particular, for later use, we note that for the MAD state $a^* \lesssim 0.59$ when $s = 0.3$, $a^* \lesssim 0.62$ when $s = 0.8$, $a^* \lesssim 0.6$ when $\Tilde{\eta} = 10/4$, $a^* \lesssim 0.6$ when $\Tilde{\eta} = 10/2$, and $a^* \lesssim 0.61$ when $\alpha = 0.1$.

\begin{figure}
    \centering
    \vspace{-5pt}
    \includegraphics[width = 1\linewidth]{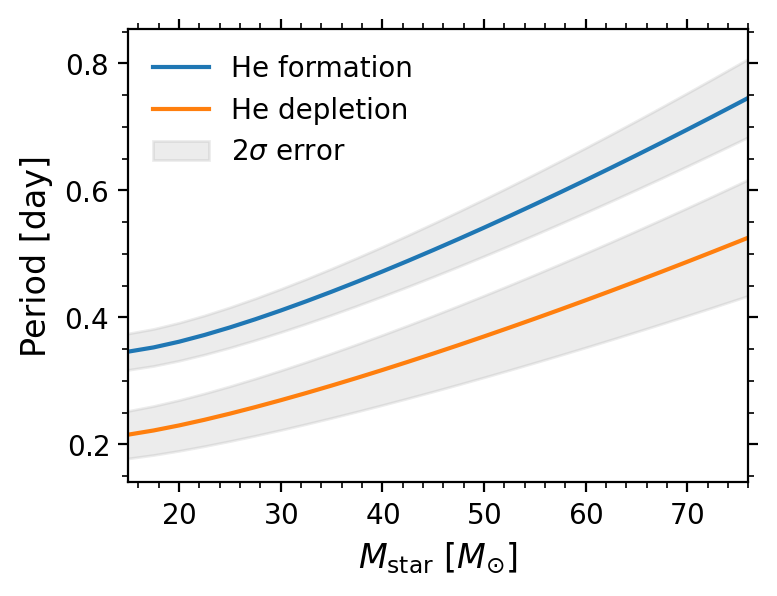}
    \vspace{-15pt}
    \caption{The maximum period at which the star would acquire and retain sufficient angular momentum to satisfy Eq.~\ref{eq: threshold AM} by core carbon depletion. The blue and orange curves assume tidal locking till the onset of the helium main sequence and core helium depletion, respectively. The x-axis shows the mass of the star at the onset of tidal locking, and the y-axis shows the period of the binary assuming that the other component is a black hole with mass ratio $q = 1$. The grey band of uncertainty results from the variance in the value of $r_g$.}
    \label{fig: domain_of_relavance}
\end{figure}

\vspace{-5pt}
\section{Discussion} \label{sec: discussion}

In this work, we have adopted two types of accretion flows that could occur near the region surrounding the black hole during collapsar dynamics to predict the black hole's mass and spin. While a magnetically arrested accretion flow can be invoked to produce the antiparallel GRB jets via the \cite{Blandford_Znajek_1977} mechanism, jets can also be produced in astrophysical objects lacking a black hole at the center (e.g., \citealt{Metzger_2011, Metzger:2017, Smith:2012}). Hence, the energy of the jets need not come from the rotational energy of the black hole. For example, one can expect the formation of anti-parallel jets purely from the energy released during the accretion flow via the \cite{Blandford_Payne:1982} mechanism. Therefore, the approach of employing a Novikov-Thorne accretion flow for the collapse dynamics is also relevant. However, a hydrodynamical flow (such as in the case of an NTD) would still require the presence of magnetic field lines to channel the outflow \citep{Blandford_Payne:1982}. In the following, we discuss some implications of the results detailed in Section~\ref{sec: results} depending on whether an NTD or MAD operates to produce the GRB jets followed by some caveats and sources of uncertainty in our work.

% \vspace{-5pt}
\subsection{Impact of stellar rotation on the lower end of PISN mass-gap} \label{sec: PISN}

Stars with helium cores in the mass range of $\approx 60 M_\odot -130 M_\odot$ may develop the right condition for $e^-e^+$ pair-production induced pair-instability supernova (PISN). This results in a (local) upper bound on the value of $M_{\rm exp}$ that could survive during core oxygen fusion \citep{Fowler_and_Hoyle1964, Rakavy_Shaviv:1967} without being completely disrupted due to the aforementioned process. Consequently, one also expects the formation of an upper mass-gap in the black hole mass spectrum. 

However, rapid rotation can shift the lower boundary of PISN to larger $M_{\rm exp}$ values \citep{Glatzel:1985}. \cite{Marchant2020_PISN_rot} found that for such systems, the lower edge of the black hole mass-gap can move upwards by $4\%-15\%$ depending on the efficiency of angular momentum transport within the progenitor stars during their evolution. In contrast, the present work suggests that if the collapse of these rapid rotators\footnote{We allow for efficient angular momentum transport within the stellar interior aided by the \cite{Spruit2002} dynamo. Turning this off would preserve more angular momentum within the star and result in favorable conditions for producing rapid rotators.} operates under the NTD formalism, such that $a^* \gtrsim 0.9$, then the maximum black hole mass for such models should converge around $M_{\rm BH} \approx 35M_\odot$ for the fiducial choice of parameters. A noticeable deviation from this value occurs when $\Tilde{\eta}$ is changed to a non-fiducial value. However, this also does not result in $M_{\rm BH}$ that are much larger that $35 M_\odot$. As the latter value is much lower than the expected lower end of the mass-gap near $M_{\rm BH} \in~\approx~[45 M_\odot, 55 M_\odot]$, e.g., \cite{Farmer_PISNe_2019, Marchant2020_PISN_rot}, consequently, rapid rotators under a Novikov-Thorne accretion flow seem unlikely to extend the lower end of the PISN mass-gap.

\vspace{-5pt}
\subsection{Case for a massive and maximally spinning stellar black hole} \label{sec: mass of maximal BH}

We find that in MAD state, the maximum black hole spin accumulates near $a^* \approx 0.62$ (see Section~\ref{subsec: Variation of the fiducial parameters}) while for in NTD state, it takes the value $a^* \approx 1$. However, for the latter, when $a^* \approx 1$, the collapsar energetic places an upper limit on the black hole mass as $M_{\rm BH} \approx 35 M_{\odot}$, which could be even smaller given that the $M_{\rm exp}$ values corresponding to these massive black holes is large enough for them to experience pair-instability supernovae, e.g., \cite{Farmer_PISNe_2019, Marchant2020_PISN_rot}. Hence, irrespective of the nature of the accretion flow, the present work suggests that a near maximally rotating black hole (i.e., $a^* \approx 1$) with $M_{\rm BH} \gtrsim 35 M_\odot$ might not result directly from stellar collapse. Instead, such a black hole could either be a second-generation black hole (i.e., a merger product of an earlier BBH system, e.g., \citealt{Pretorius:2005, Fishbach:2017, Gerosa:2017}) or a product of subsequent mass (and angular momentum) accretion (e.g., \citealt{Gammie:2004, Sadowski_2011}). For the latter, super-Eddington accretion would be the likely scenario, as sub-Eddington accretion would make it difficult to spin up black holes to critical within the lifetime of a massive mass-donor star. Cygnus X-1 with $a^* > 0.95$ (\citealt{CygnusX1:2011}; assuming that most of $a^*$ is acquired from the direct collapse of its progenitor star) may appear as a direct counterexample to the above statement, namely '' black holes with $a^* \approx 1$ and $M_{\rm BH} \gtrsim 35 M_\odot$ might not result directly from stellar collapse''. However, we note that Cygnus X-1 has a mass of $\approx 21 \, M_\odot$, which falls below the derived mass limit.

% \vspace{-5pt}
\subsection{A possible upper bound on \texorpdfstring{$\chi_{\rm eff}$}{} in black hole - helium star binaries} \label{sec: max val of chi_eff}

From Fig.~\ref{fig: 2D_hist_spin}, we see that for a given $M_{\rm exp}$, a higher value of $\ell_{\rm exp}$ does not always lead to a larger $a^*$. Instead, once Eq.~\ref{eq: threshold AM} is satisfied, the assumption of the nature of the accretion flow becomes crucial. For the case of MAD, a rapid spin-down is observed, resulting in $a^* \approx 0.12$ (cf. \citealt{Gottlieb:2023, Jacquemin_Ide:2023}). This results in an upper bound on the spin with $a^* \lesssim 0.62$. In the current 1D approach, the latter value is also not very sensitive to a change in the magnitude of $s, \alpha$ or $\Tilde{\eta}$, (e.g., see Section~\ref{subsec: Variation of the fiducial parameters}).
In such a case, under the scenario when only the second-born black hole acquires spin in a tidally locked helium star-black hole binary (e.g., see Section~\ref{sec: spin up via tidal interaction}), an upper bound on $\chi_{\rm eff}$ can be estimated for such systems as
\begin{equation}
    \chi_{\rm eff, max} = \frac{a^*_2}{q^{-1}_{\rm max} + 1} \approx 0.31 \,,
\end{equation}
where $q:= M_2/M_1 \leq 1$ and we set $q_{\rm max} = 1$. 

On the other hand, the LIGO-Virgo-KAGRA collaboration \citep{Abbott_O3a_2021, Abbott_O3b_2021} has identified more than half a dozen candidate BBHs that do not respect this equality (see, e.g., Table~\ref{tab: LVK inferred paramters}). If MAD is the mechanism behind GRBs, this hints that an alternative scenario might be behind the formation of such BBHs. Tidal CHE can produce systems with large $\chi_{\rm eff}$ values since, in this case, both black holes can acquire large spin. However, the latter requires nearly equal mass-ratio binaries \citep{Marchant2016}, which is not the case for the black holes in Table~\ref{tab: LVK inferred paramters}.  Alternatively, super-Eddington accretion onto the first formed black hole in an ordinary binary system can also generate a rapidly spinning (and massive) black hole. However, the systems in Table~\ref{tab: LVK inferred paramters} likely result from massive stellar progenitors that have a lifespan of only a few million years. This allows a tiny window for mass transfer, which implies the requirement of a rapid mass-transfer rate supported by a very strong super-Eddington accretion. Additionally, the period of the binary should be such that they could subsequently merge within a few billion years. It is uncertain as to how likely these conditions could be met simultaneously.

A more plausible formation pathway for BBHs in Table~\ref{tab: LVK inferred paramters} could be via hierarchical merger (e.g., \citealt{Fishbach:2017, Gerosa:2017, Kimball:2021}), e.g., in a triple system or dynamical capture in stellar clusters or in the disk of active galactic nuclei (AGN, e.g., \citealt{Gerosa:2021, Sedda:2023} and reference therein). This is because the merger of even non-rotating black holes can result in second-generation black holes with $a^* \approx 0.7$ (e.g., \citealt{Pretorius:2005}). Additionally, BBHs living in the disks of AGN also have the possibility of accreting mass and angular momentum from the disk.

% \vspace{-5pt}

% The primary difference lies in the orbital dynamics. The Kepleiarn orbits of Newtonian dynamics are not the same in Kerr geometry. Frame dragging is absent too. However, both these aspects do not affect the black hole property in our time-averaged dynamics approach. 

% \vspace{-5pt}
\subsection{Caveats and uncertainties}

\begin{enumerate}

    \item The final structure of the star has been deemed crucial for the following explosion, with some massive stars producing a neutron star instead of a black hole, leading to the so-called islands of explodeability \citep{OConnor2011black, Sukhbold_2016, sukhbold2018high}. Although it remains to be seen how this fares in the case of rapidly rotating stars where the dynamics of core collapse are dominated by the collapsar energetics. However, there is a possibility for a magnetar-like explosion \citep{Woosley_2010, Kasen_Bildsten_2010} where the energetics are dictated by the magnetic fields of a proto-neutron star instead of a black hole. We ignore such a scenario in this work. As such, in our analysis, a black hole always forms during the collapse of the core of a massive star.

    \item   At times, the stochasticity of the core structure can influence the nature of remnant masses. However, any stochasticity would be lost in our approach due to the averaging performed in the analysis. As discussed in \cite{Fryer:2021_remnant_mass_dist}, the effect of stochasticity would nevertheless average out for a large enough sample size, which is always true in the context of population studies. This would also be the case with observations in the coming year as the number of gravitational wave detections grows.

    \item In the case of non/slowly rotating stars, our formalism results in a direct collapse. Although this might not necessarily be the case, the lack of detections of supernova explosions from high-mass stars \citep{Smartt:2015} as well as the theoretical findings that more massive stars might disappear without leaving any observational feature \citep{Murguia_Berthier:2020}, suggest that black holes forming from more massive stars might experience direct collapse (e.g., see \citealt{Adams:2017} for a failed supernova candidate). However, we note that stochastically varying angular momentum accretion on the proto-neutron star may form jittering jets \citep{Papish_Soker_2011, Soker:2023}, which could prevent direct collapse.

    \item For the case of an NTD, the accretion flow remains circular with the disk truncating at the ISCO.  However, numerical studies find that at a large accretion rate, the inner edge of the disk moves closer to the black hole but cannot be uniquely defined \citep{Abramowicz:2010}. The flow also deviates from a Keplerian profile \citep{Abramowicz:2010, Sadowski_2011}, becoming super-Keplerain for our choice of $\alpha = 0.01$. We do not consider these aspects here. For rapid rotators, the ISCO will approach very close to the black hole, hence reducing our margin of error.

    \item In the current work, we ignore the effect of rotation deformation of the star on the mass accretion rate which may affect our results. Additionally, In the case when a non-trivial change in stellar physics results in a substantial change in the star's angular momentum profile (e.g., shown in Fig. \ref{fig: angular_momentum_profile}) at collapse, our calculations could be affected.

    \item Finally, we note that the current results are a 1D approximation of a complex 3D phenomenon. For greater precision, one would need to conduct a full 3D GRMHD study.

\end{enumerate}

\vspace{-10pt}
\section{Conclusion} \label{sec: conclusion}

Black holes born from rapidly rotating progenitor stars can experience a very different supernova explosion than those born from non/slowly rotating stars. This is because a rapidly rotating stars contain a large reservoir of angular momentum that they would need to lose to collapse below a certain characteristic radius.  Here, we studied the resulting masses and spins of the black hole when dealing with the collapse of such rotating stars. We showed that for rapid rotators, the black hole mass and spin is a function of the assumption on the accretion flow and can be significantly different from that expected for a non-/slowly rotating progenitor star (Section~\ref{sec: results}). 
This discrepancy could influence the merger time, luminosity distance, and gravitational wave properties of the BBH.  We find that rapid rotators undergoing a collapsar explosion via a Novikov-Thorne accretion flow might not extend the lower edge of the black hole upper mass-gap (Section~\ref{sec: PISN}) and that a maximally rotating black hole born directly from stellar collapse might have $M_{\rm BH} \lesssim 35M_\odot$ as its maximum possible mass (Section~\ref{sec: mass of maximal BH}). For the case of a magnetically arrested accretion flow, we find a maximum black hole dimensionless spin of $a^* \approx 0.62$. For black holes born in helium star black hole binaries, this puts an upper limit on the effective spin of the BBH as $\chi_{\rm eff} \lesssim 0.31$ (Section~\ref{sec: max val of chi_eff}).

\vspace{-15pt}
\section*{Acknowledgements}

The authors are grateful to Dr. Pablo Marchant and the anonymous referee whose feedback significantly improved this paper. SG is supported by the University of Auckland Doctoral Scholarship. JJE acknowledges support of Marsden Fund Council managed through Royal Society Te Apārangi. This work utilized NeSI high performance computing facilities.

%%%%%%%%%%%%%%%%%%%%%%%%%%%%%%%%%%%%%%%%%%%%%%%%%%

\vspace{-15pt}
\section*{Data Availability}

The numerical work discussed in Section~\ref{sec: numerical methods} can be found at \url{https://github.com/SohanGhodla/Collapsar-Formalism}. The \textsc{Sage} manifolds notebook used to conduct part of the calculation can also be found at the same URL. The simulated data on which the fits were applied have been included in the supplementary material.

%%%%%%%%%%%%%%%%%%%%%%%%%%%%%%%%%%%%%%%%%%%%%%%%%%



\vspace{-15pt}
\bibliographystyle{mnras}
\bibliography{refs} % if your bibtex file is called example.bib

\appendix

\vspace{-15pt}
\section{Freefall timescale} \label{sec: free_fall time}

For a shell made of perfect fluid, the collapse is such that the coordinate $\theta$ remains unchanged. This means that the trajectory of the particles would not cross each other during infall. Solving for the geodesic motion (assuming an infall from rest at radial infinity), one can show that for such a case 
\begin{equation}
    \frac{dr}{dt} = - \frac{(r^2 - 2Mr + a^2) [2Mr(r^2 + a^2)]^{1/2} }{ (r^2 + a^2)^2 - (r^2 - 2Mr + a^2) a^2 \sin^2 {\theta}} \,.
    \label{eq: infall coordinate time}
\end{equation}
Eq.~\ref{eq: infall coordinate time} depends on the coordinate $\theta$, implying that in a Kerr geometry, the equatorial part of a spherical shell falls faster than the polar region (e.g., \citealt{bivcak1976fall}).  
Integrating Eq.~\ref{eq: infall coordinate time} when $a \neq 0$ is difficult, and the resulting level of precision is not required here. For example, in Fig.~\ref{fig: infall coordinate time}, we compare the infall coordinate time (resulting from integrating Eq.~\ref{eq: infall coordinate time} with $a=0$) with the infall proper time, assuming that both time coordinates are synchronized at some fixed distance $r_0 =30M$ (which we approximate as the minimum distance from where an infalling shell would first circularize at ISCO in our work). 
We find that upon reaching ISCO, there is $\approx 12\%$ disagreement in the two time intervals. As $r_0$ moves to larger values, this disagreement becomes smaller. E.g., for $r_0 = 200M$, it is reduced to $\approx 2\%$. Finally, when calculating the disk formation time $\Delta t$ in Section~\ref{sec: formation of the disk}, we note that the matter falls from a finite radius $r_0$, from a near-rest configuration, which also needs to be taken into account.
% 
% \begin{equation}
%     \Delta t = - \left[ {2 M \ln\left(\frac{\sqrt{r} - \sqrt{2M}}{\sqrt{r} + \sqrt{2M}}\right) + 2\sqrt{2Mr} + \frac{3  a^{2} \sqrt{\frac{2M}{r}} + 2\sqrt{2M r^3}}{6M}} \right]_{r_{\rm isco}}^{r_0} .
% \end{equation}
% 
\begin{figure}
    \centering
    % \vspace{-10pt}
    \includegraphics[width = 0.9\linewidth]{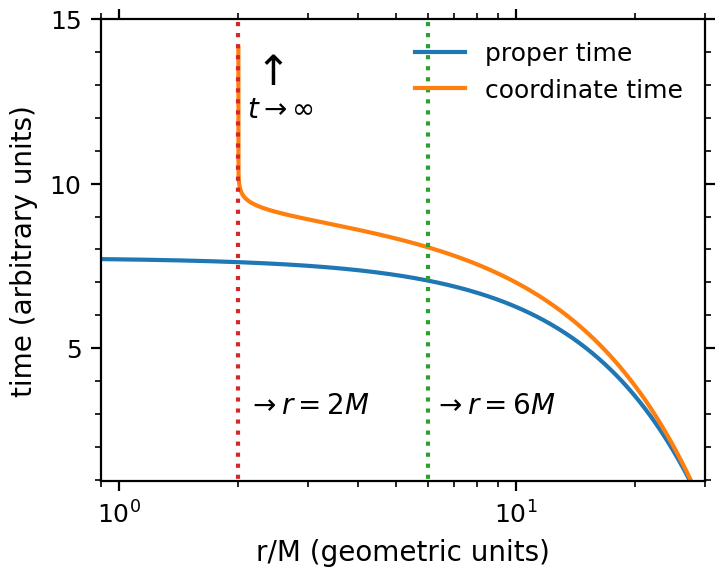}
    % \vspace{-5pt}
    \caption{The deviation in the infall coordinate time and the infall proper time (in a Schwarzschild geometry), assuming that both time coordinates are synchronized at some fixed distance $r_0 =30M$, which we approximate as the minimum distance from where an infalling shell would first circularize at ISCO ($r=6M$) in our work.}
    \label{fig: infall coordinate time}
\end{figure}

% \vspace{-15pt}
\section{An estimate for $\theta_c$} \label{sec: estimating theta_c}

Let us define the breakout time of the jet from the stellar surface as $t_{\mathrm{br}} = \Tilde{\eta} r_* / c$, where $ \Tilde{\eta} > 1$ and is a measure of how fast the jet head travels through the stellar material (i.e., $\Tilde{\eta}$ is the ratio of the speed of light to the jets' head speed). Then, the relativistic cocoon half-opening angle can be defined as

\begin{equation}
    \theta_c \approx \frac{r_{\perp}}{r_*} \approx  \frac{v_{\rm sh}  
 t_{\mathrm{br}}}{r_*} \,.
\end{equation}
Above $r_*, r_\perp$ is the radius of the star and the transverse radius of the cocoon and $v_{\rm sh}$ is the velocity of the shock wave in the transverse direction and takes the form $v_{\rm sh} = \sqrt{ \frac{p_c} {\rho_*}}$ \citep{Lazzati:2005}, where $p_c, \rho_*$ are the pressure within the cocoon and the mean density of the star. \cite{Lazzati:2005} calculates the pressure as

\begin{equation}
    p_c \approx \left(\frac{L_j \rho_*}{3 r_* t_{\mathrm{br}}}\right)^{1 / 2} \,,
\end{equation}
where $L_j$ is the luminosity of the jet. This gives
\begin{equation}
    \theta_c \approx  \Tilde{\eta}^{3/4} \left(\frac{L_j}{3 r_*^2 \rho_* c^3}\right)^{1 / 4} \,.
\end{equation}
Unfortunately, $ \Tilde{\eta}$ remains a free parameter in the present work. However, the jet head moves slowly $(\approx 0.3c)$ within the baryonic medium of the star \citep{Nakar:Piran:2017}, implying that $ \Tilde{\eta} \approx 10/3$.

\vspace{-10pt}
\section{Additional Material} \label{sec: variation of fiducial parameters}

% \vspace{-15pt}

\begin{figure}
    \centering
    % \vspace{7pt}
    \includegraphics[width = 1\linewidth]{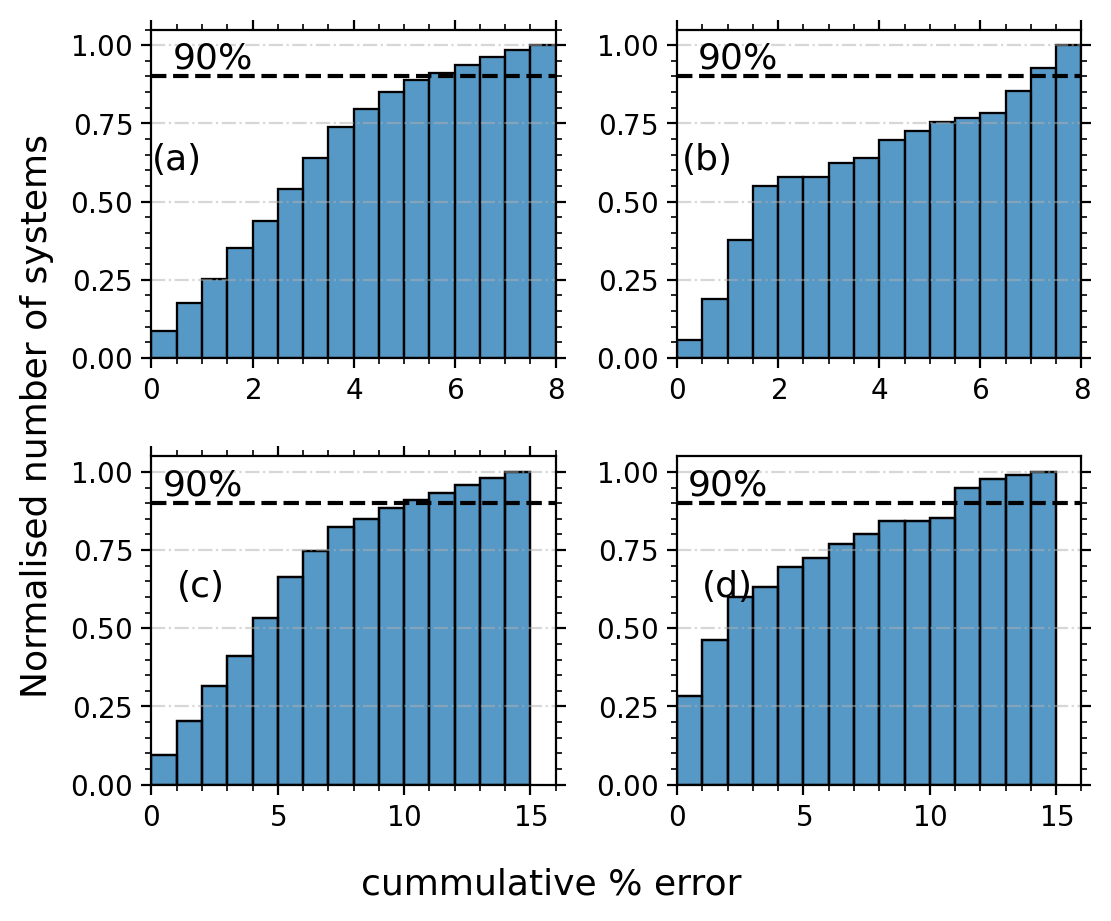}
    % \vspace{-5pt}
    \caption{Figure (a) and (b) show the fitting error in the black hole masses given in Eq.~\ref{eq: 2d_MBH_fit}, where Figure (a) is for NTD and Figure (b) for MAD state. Figure (c) and (d) show the fitting error in the black hole spins given in Eq.~\ref{eq: 2d_spin_fit}, where Figure (c) is for NTD and Figure (d) for MAD state. The intersection of the histogram with the horizonal line shows the percentage error within which $90\%$ of the systems tend to lie - as predicted by Eq.~\ref{eq: 2d_MBH_fit} or Eq.~\ref{eq: 2d_spin_fit} - when compared to the true value.}
    \label{fig: error_MBH}
\end{figure}

% \vspace{-15pt}
% \input{table2.tex}
\begin{table}
% \vspace{-10pt}
    \centering
    \caption{LVK inferred parameters for systems with $\chi_{\rm eff} \geq 0.32$ \citep{Abbott_O3a_2021, Abbott_O3b_2021}}
    \label{tab: LVK inferred paramters}
    
    \begin{tabular}{cccc} % four columns, alignment for each
    \hline

    Name & $M_1 \, [M_\odot]$  & $M_2 \, [M_\odot]$  & $\chi_{\rm eff}$ \\
    \hline
    \hline
     
    GW190517\_055101  & $39.2^{+13.9}_{-9.2}$ & $24.0^{+7.4}_{-7.9}$ & $0.49^{+0.21}_{-0.28}$  \\
    GW200208\_222617 & $51^{+103}_{-30}$ & $12.3^{+9.2}_{-5.5}$ &$ 0.45^{+0.42}_{-0.46}$  \\
    GW170729          & $50.2^{+16.2}_{-10.2}$ & $34.0^{+9.1}_{-10.1}$ & $0.37^{+0.21}_{-0.25}$  \\
    GW190805\_211137  & $46.2^{+15.4}_{-11.2}$ & $30.6^{+11.8}_{-11.3}$ & $0.37^{+0.29}_{-0.39}$  \\
    GW190620\_030421  & $58.0^{+19.2}_{-13.3}$ & $35.0^{+13.1}_{-14.5}$ & $0.34^{+0.22}_{-0.29}$  \\
    GW190519\_153544  & $65.1^{+10.8}_{-11}$ & $40.8^{+11.5}_{-12.7}$ & $0.33^{+0.20}_{-0.24}$  \\
    GW200306\_093714  & $28.3^{+17.1}_{-7.7}$ & $14.8^{+6.5}_{-6.4}$ & $0.32^{+0.28}_{-0.46}$  \\
    \hline
    \end{tabular}
\end{table}

% ------------------------------------
\begin{figure*}
    \centering
    \vspace{-5pt}
    \includegraphics[width = 0.95\linewidth]{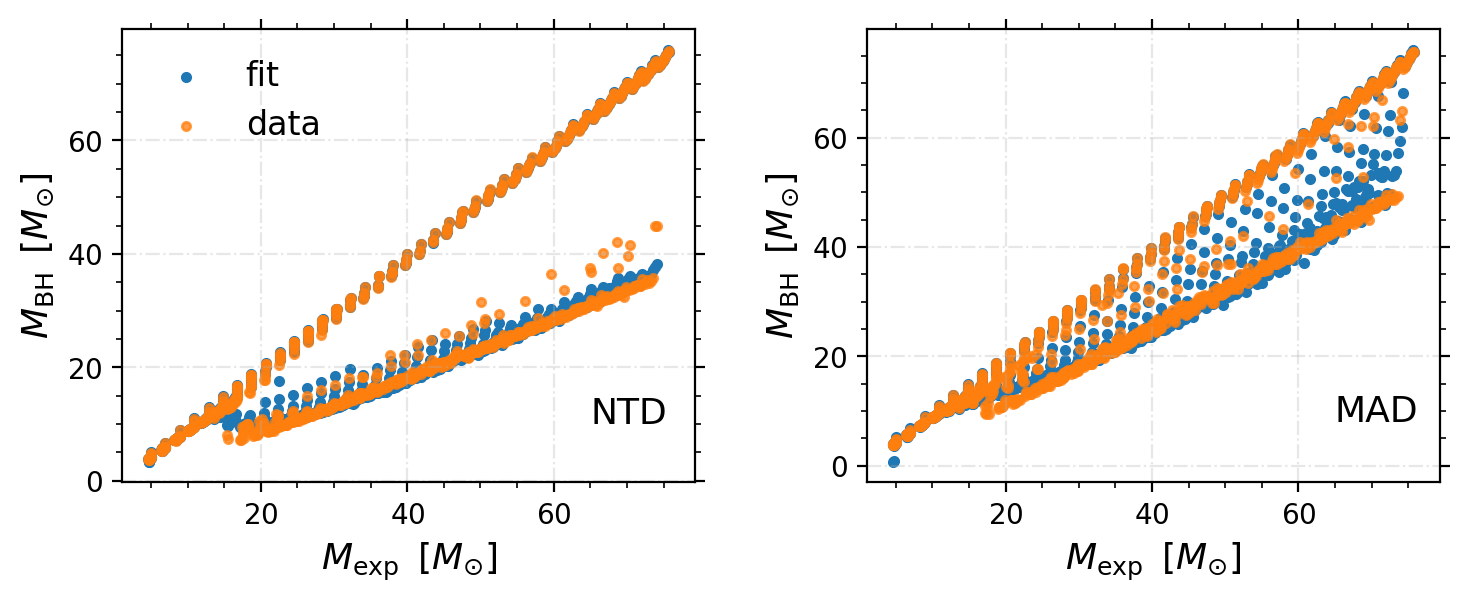}
    \vspace{-5pt}
    \caption{A comparison between the fit predictions (blue dots) and the simulation data (orange dots) for the black hole mass in the NTD and the MAD state.}
    \label{fig: MBH_error_comparison}
\end{figure*}

\begin{figure*}
    \centering
    \vspace{-5pt}
    \includegraphics[width = 0.95\linewidth]{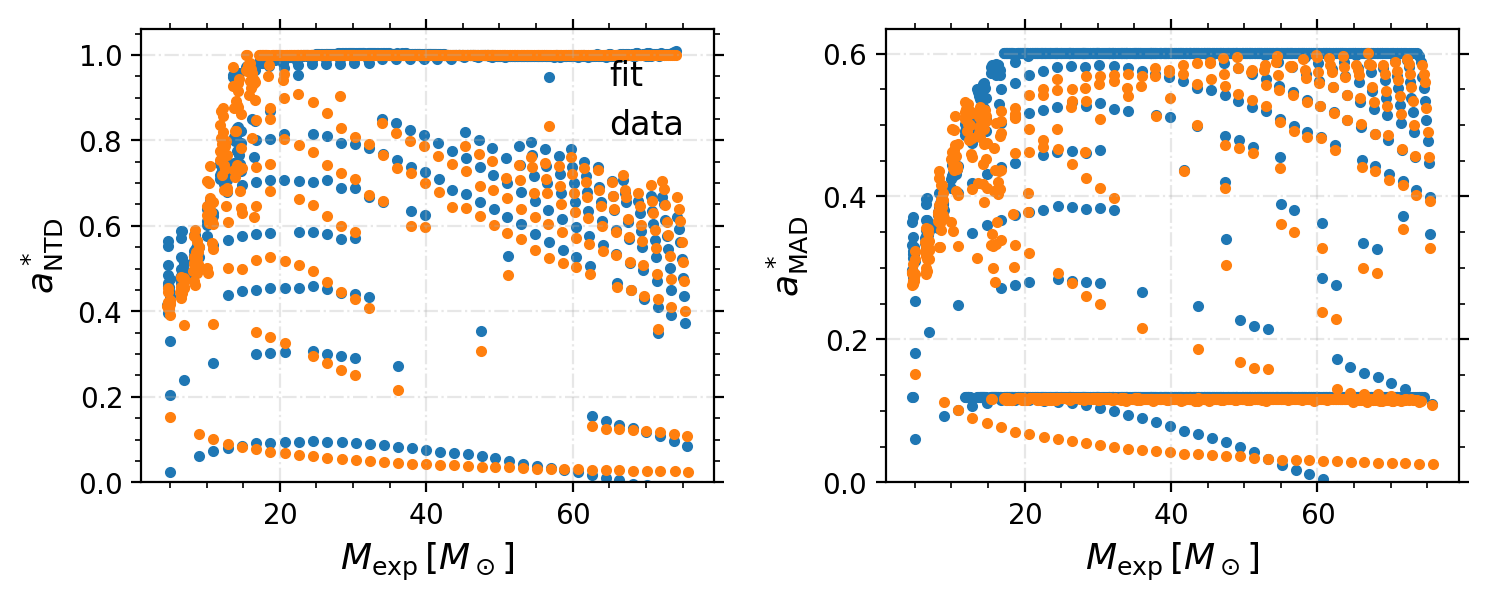}
    \vspace{-5pt}
    \caption{A comparison between the fit predictions (blue dots) and the simulation data (orange dots) for the black hole spin in NTD and the MAD state.}
    \label{fig: spin_error_comparison}
\end{figure*}
% ------------------------------------

\begin{figure*}
    \centering
    \vspace{-5pt}
    \includegraphics[width = 0.95\linewidth]{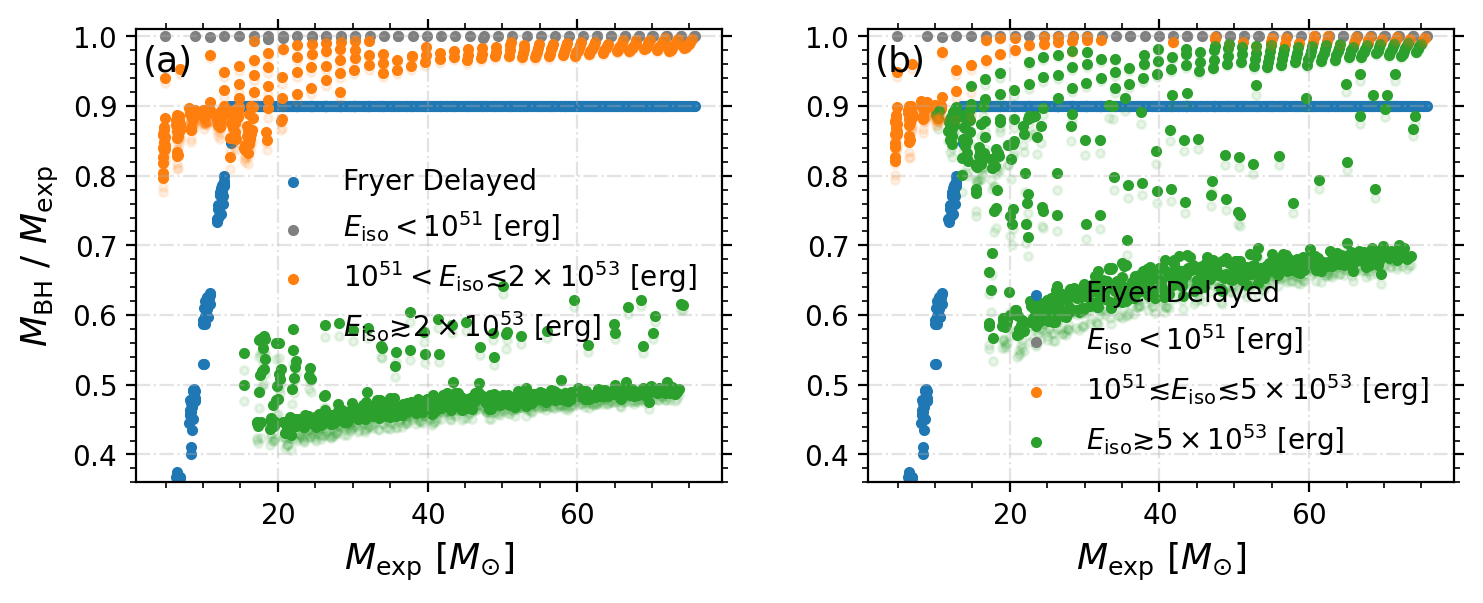}
    \vspace{-5pt}
    \caption{The figure is the same as Fig.~\ref{fig: Mexp_vs_MBH_ratio}. Figure (a) is for the NTD state, while Figure (b) is for the MAD state. The dark color is for the case when $s = 0.3$, while the light color for $s = 0.5$. The latter is the same as Fig.~\ref{fig: Mexp_vs_MBH_ratio}.}
    \label{fig: spt3}
\end{figure*}

\begin{figure*}
    \centering
    \vspace{-5pt}
    \includegraphics[width = 0.95\linewidth]{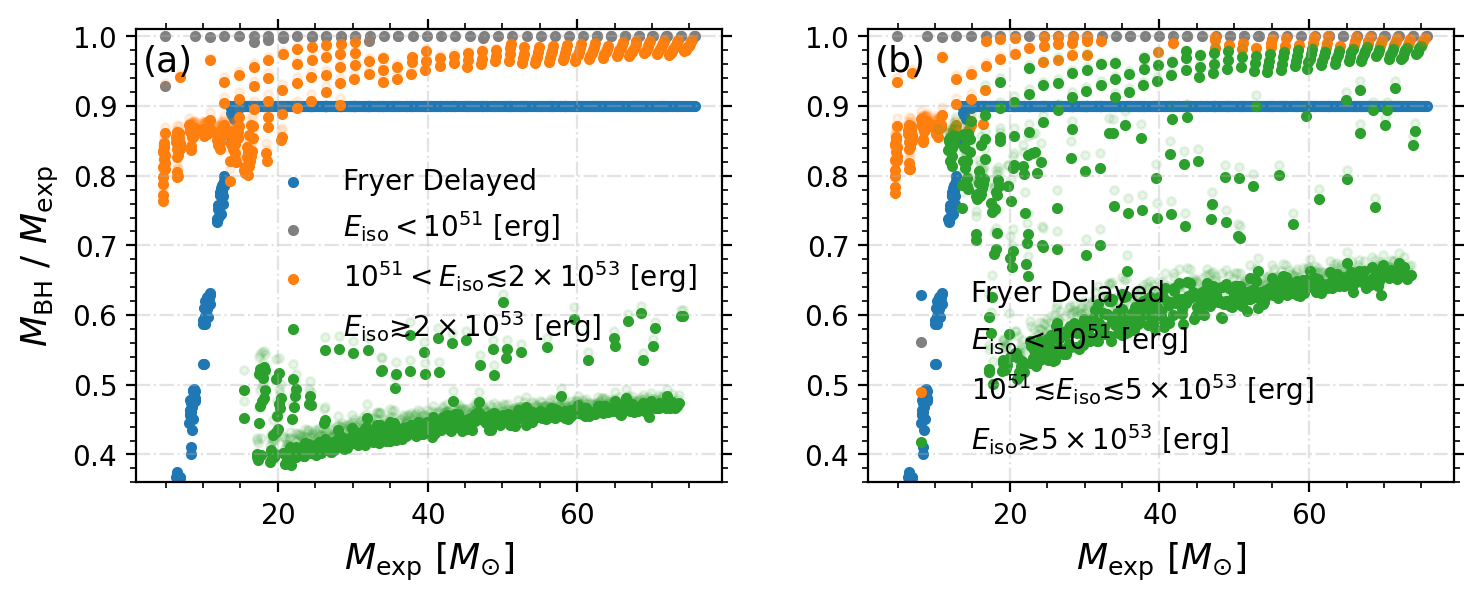}
    \vspace{-5pt}
    \caption{Same as Fig.~\ref{fig: spt3} but with $s = 0.8$.}
    \label{fig: spt8}
\end{figure*}

\begin{figure*}
    \centering
    \vspace{-5pt}
    \includegraphics[width = 0.95\linewidth]{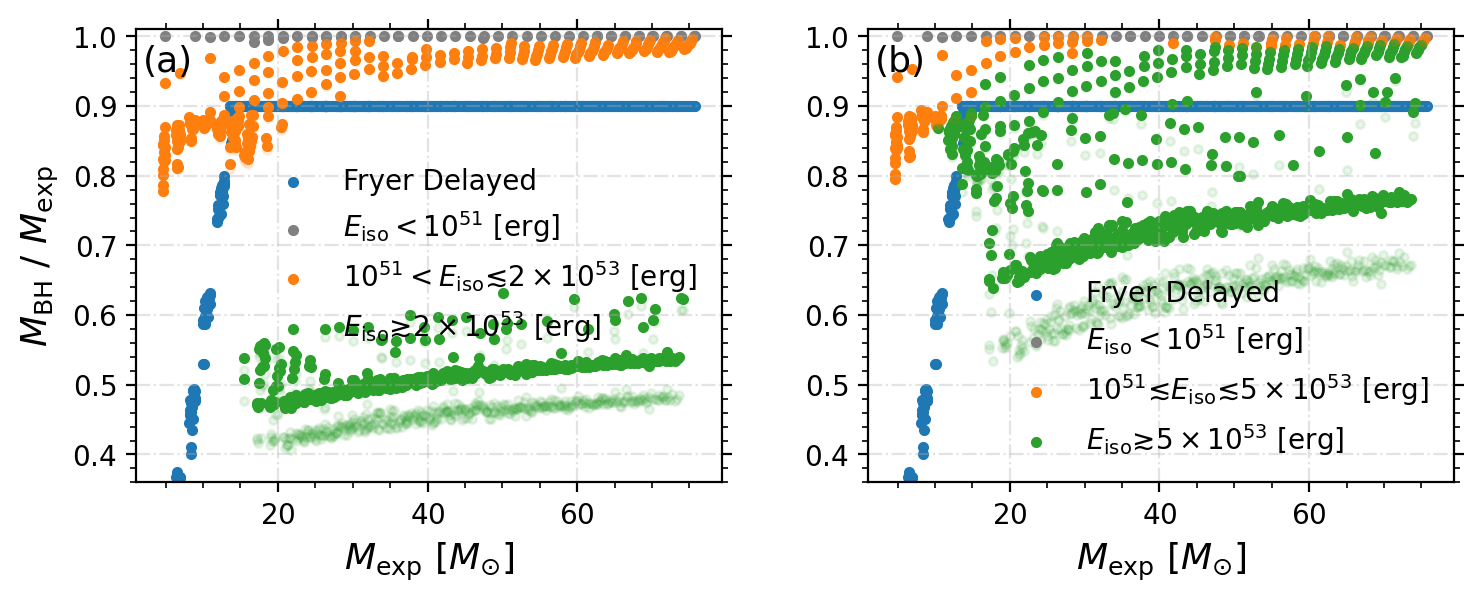}
    \vspace{-5pt}
    \caption{Same as Fig.~\ref{fig: spt3} but with instead of changing $s$ here we have changed $\Tilde{\eta}$.  The dark color is for the case when $\Tilde{\eta} = 10/4$., while the light color for the fiducial case ($\Tilde{\eta} = 10/3$.). The latter is the same as Fig.~\ref{fig: Mexp_vs_MBH_ratio}.}
    \label{fig: eta_2pt5}
\end{figure*}

\begin{figure*}
    \centering
    \vspace{-5pt}
    \includegraphics[width = 0.95\linewidth]{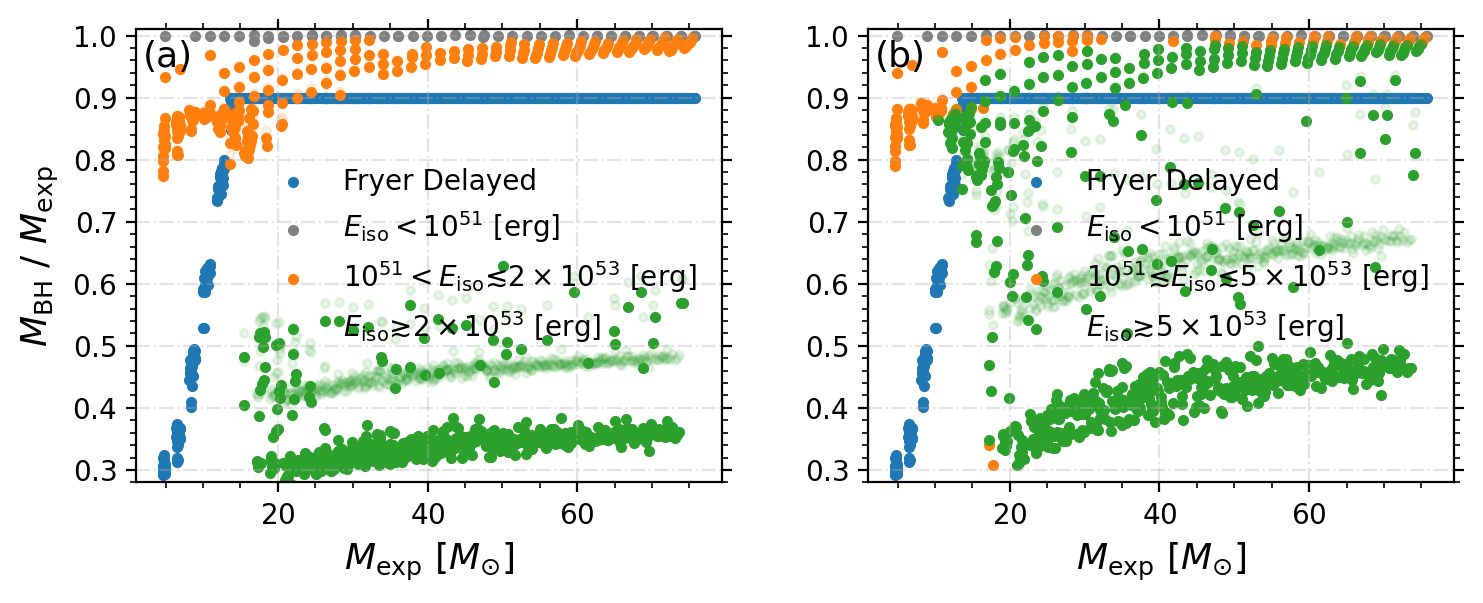}
    \vspace{-5pt}
    \caption{Same as Fig.~\ref{fig: eta_2pt5} but with $\Tilde{\eta} = 10/2$. Note the change in the lower range of the y-axis.}
    \label{fig: eta_5}
\end{figure*}

\begin{figure*}
    \centering
    \vspace{-5pt}
    \includegraphics[width = 0.95\linewidth]{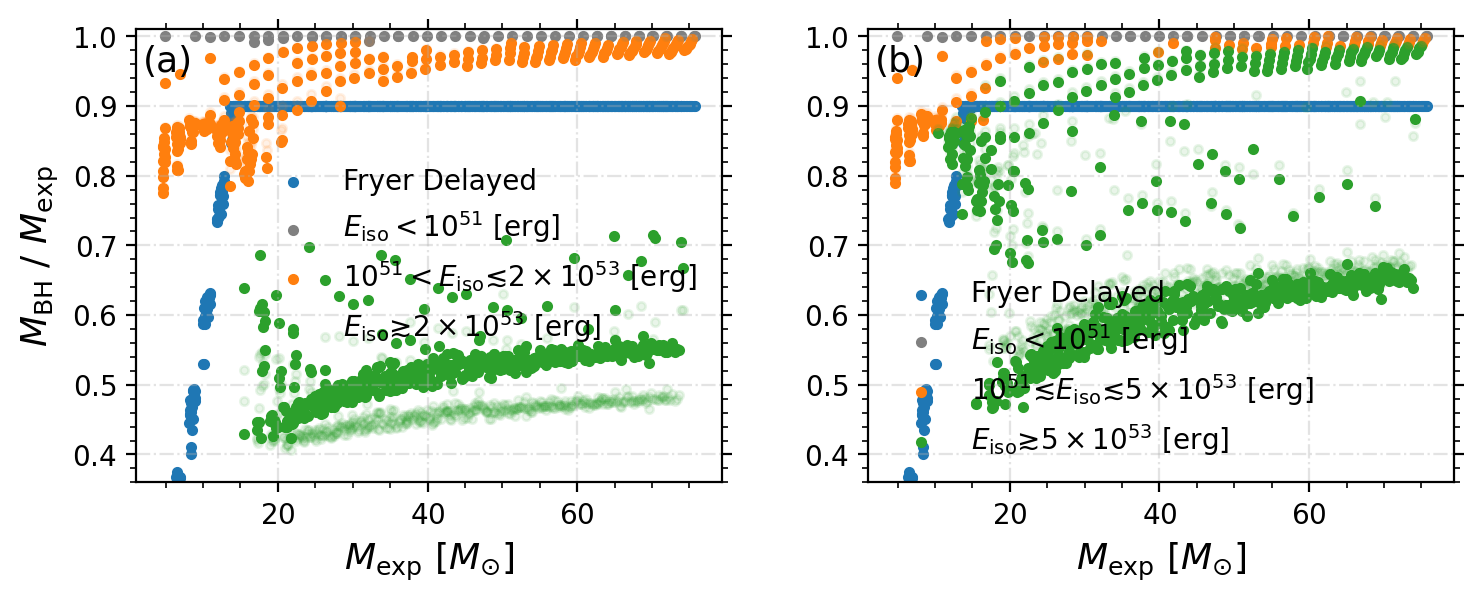}
    \vspace{-5pt}
    \caption{Same as Fig.~\ref{fig: spt3} but instead of changing $s$ here we have changed $\alpha$.  The dark color is for the case where $\alpha = 0.1$, while the light color is for $\alpha = 0.01$. The latter is the same as Fig.~\ref{fig: Mexp_vs_MBH_ratio}.}
    \label{fig: alpha_pt1}
\end{figure*}

\begin{figure*}
    \centering
    \vspace{-5pt}
    \includegraphics[width = 1\linewidth]{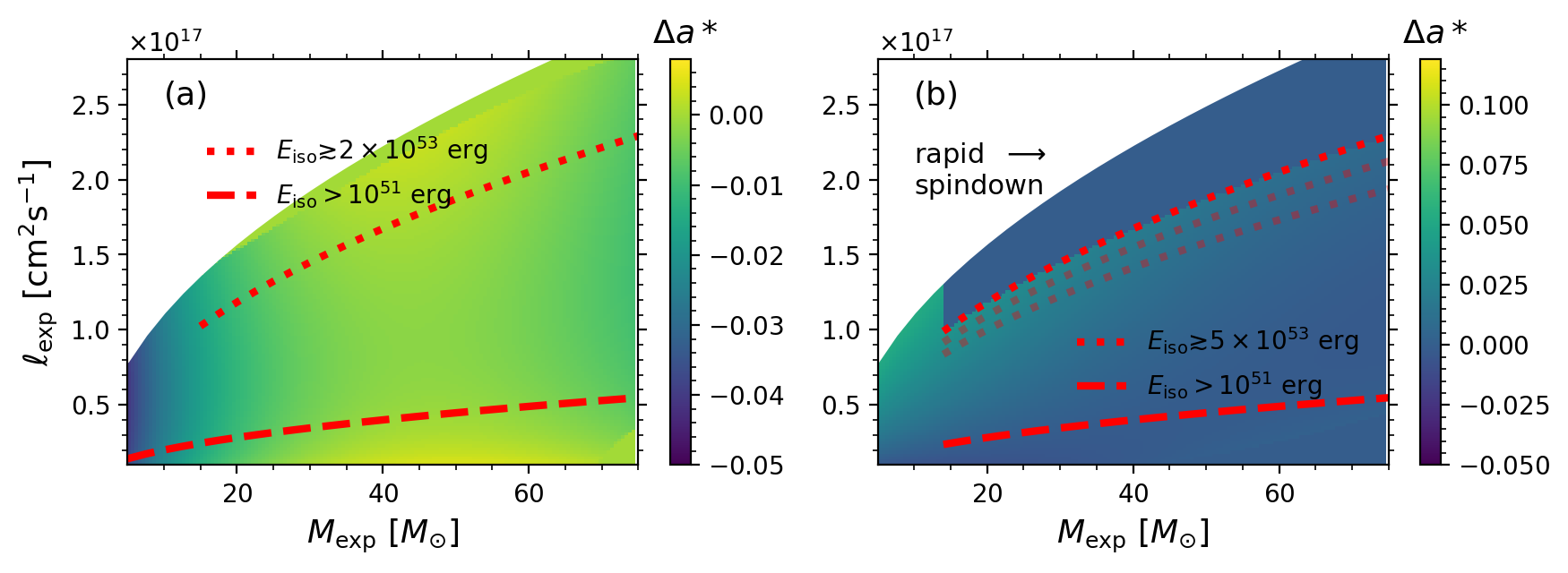}
    \vspace{-5pt}
    \caption{Same as Fig.~\ref{fig: 2D_hist_spin} but with $s = 0.3$. The colorbar now shows the \textit{residual value} of $a^*$, i.e. $ \Delta a^* = a^*_{\rm fiducial} - a^*_{s = 0.3}$.}
    \label{fig: spin_spt3}
\end{figure*}

\begin{figure*}
    \centering
    \vspace{-5pt}
    \includegraphics[width = 1\linewidth]{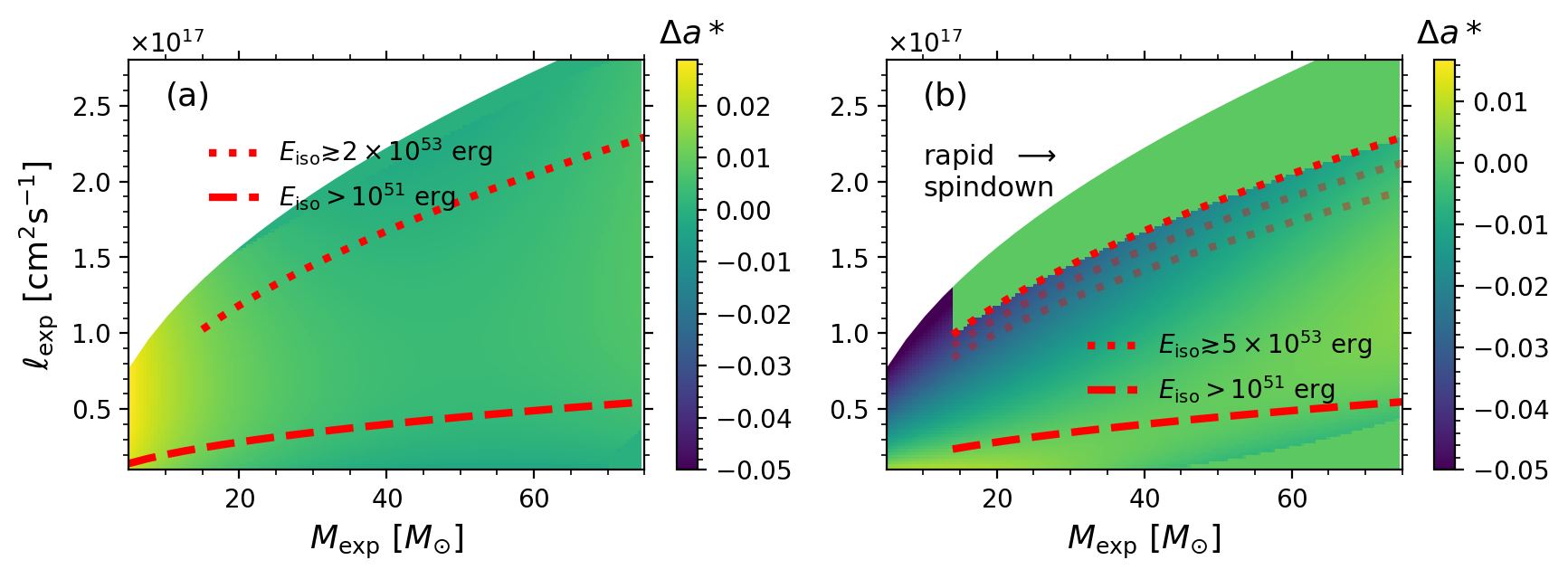}
    \vspace{-5pt}
    \caption{Same as Fig.~\ref{fig: spin_spt3} but with $s = 0.8$.}
    \label{fig: spin_spt8}
\end{figure*}

\begin{figure*}
    \centering
    \vspace{-5pt}
    \includegraphics[width = 1\linewidth]{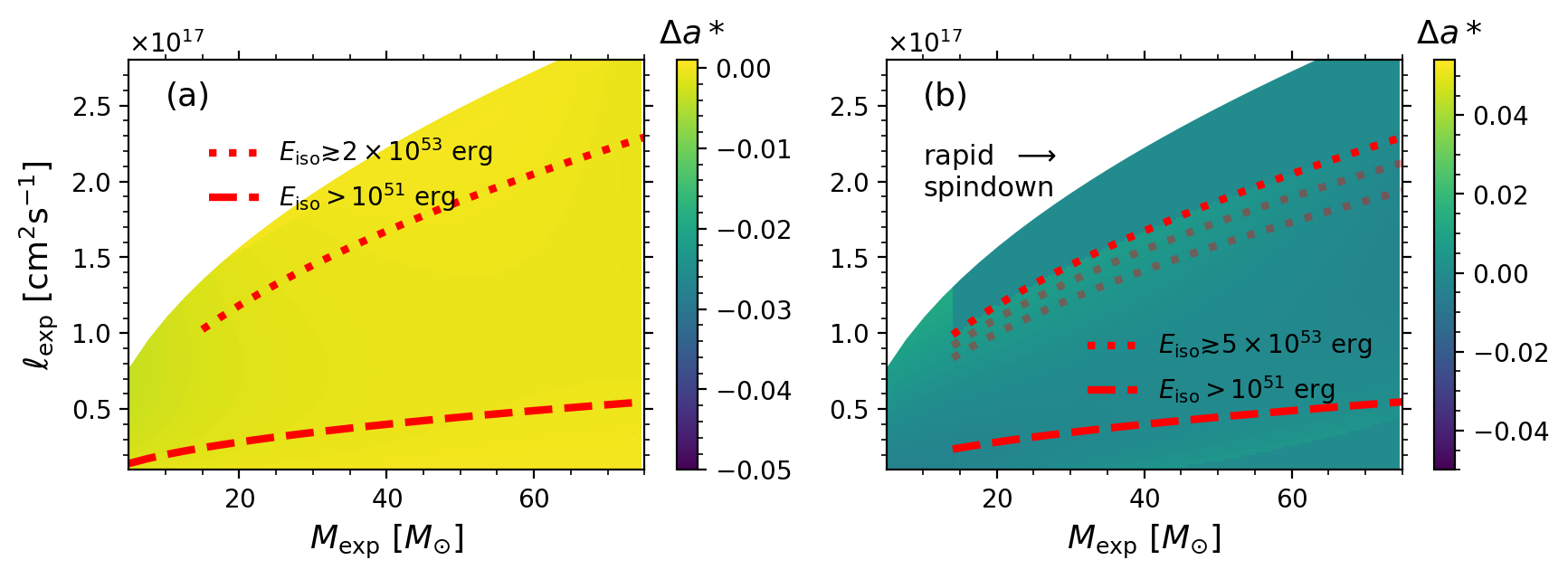}
    \vspace{-5pt}
    \caption{Same as Fig.~\ref{fig: spin_spt3} but instead of changing $s$ here we change $\Tilde{\eta}$ to $\Tilde{\eta} = 10/4$.}
    \label{fig: spin_eta_2pt5}
\end{figure*}

\begin{figure*}
    \centering
    \vspace{-5pt}
    \includegraphics[width = 1\linewidth]{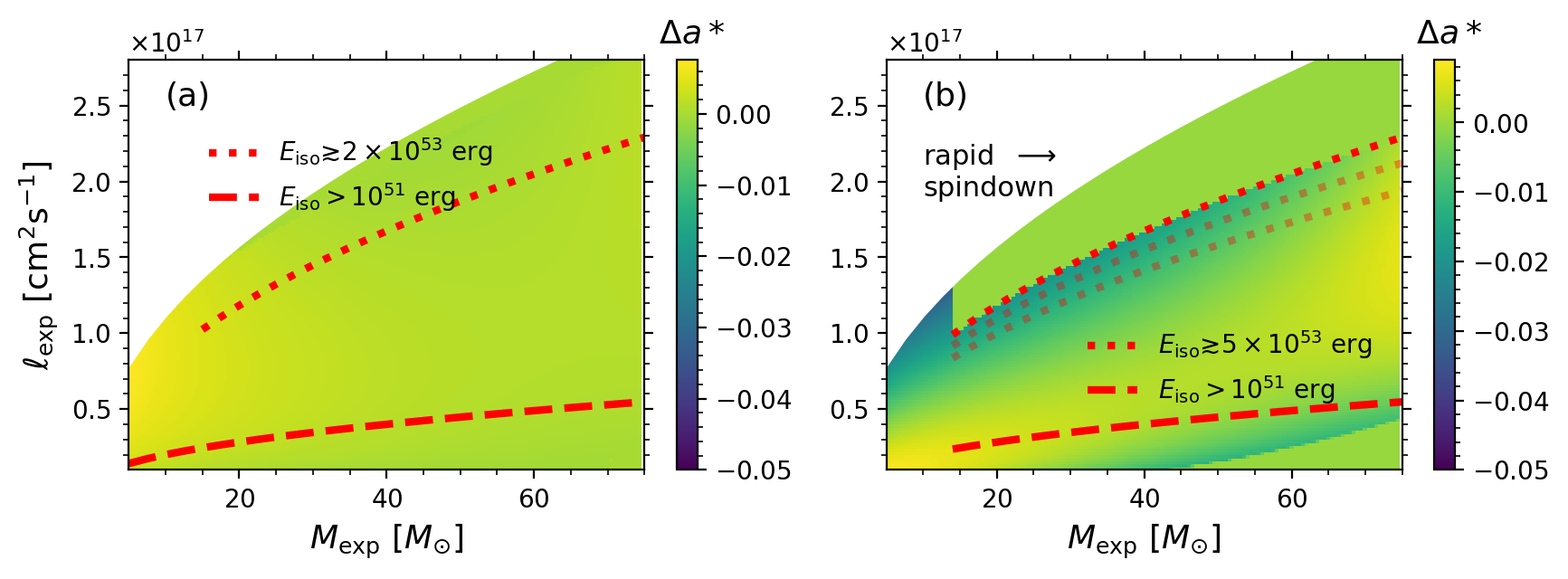}
    \vspace{-5pt}
    \caption{Same as Fig.~\ref{fig: spin_eta_2pt5} but with $\Tilde{\eta} = 10/2$.}
    \label{fig: spin_eta_5}
\end{figure*}

\begin{figure*}
    \centering
    \vspace{-5pt}
    \includegraphics[width = 1\linewidth]{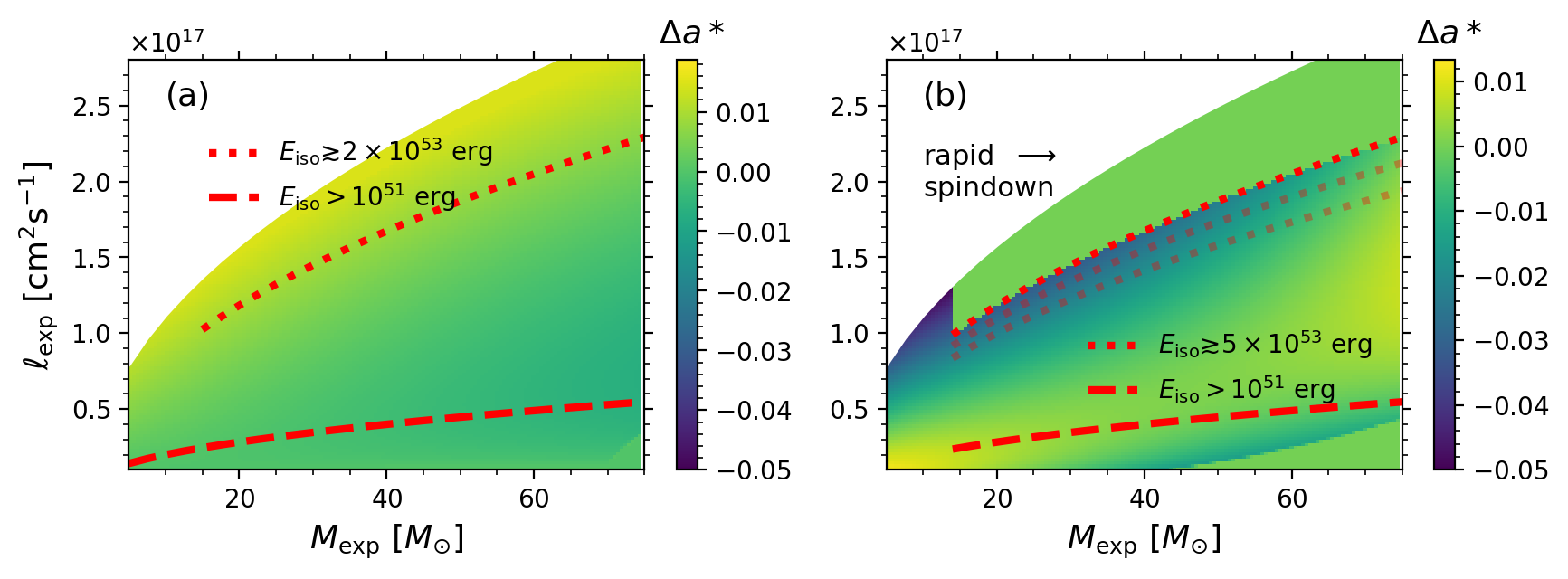}
    \vspace{-5pt}
    \caption{Same as Fig.~\ref{fig: spin_spt3} but instead of changing $s$ here we change $\alpha$ to $\alpha = 0.1$.}
    \label{fig: spin_alpha_pt1}
\end{figure*}

%%%%%%%%%%%%%%%%%%%%%%%%%%%%%%%%%%%%%%%%%%%%%%%%%%

% Don't change these lines
\bsp	% typesetting comment
\label{lastpage}
\end{document}